\documentclass{emulateapj}
\usepackage{natbib}

\def\etal{{et al.~}}
\shorttitle{SNR LF}
\shortauthors{Chomiuk \etal}

\begin{document}

\title{A Universal Luminosity Function for Radio Supernova Remnants}
\author{Laura Chomiuk\altaffilmark{1} \& Eric M. Wilcots\altaffilmark{1}}
\email{chomiuk@astro.wisc.edu}
\altaffiltext{1}{University of Wisconsin--Madison, Madison, WI 53706}
\begin{abstract}
We compile radio supernova remnant (SNR) samples from the literature for 19 nearby galaxies ranging from the SMC to Arp 220, and use this data to constrain the SNR luminosity function (LF) at 20 cm. We find that radio SNR populations are strikingly similar across galaxies. The LF can be described as a power law with constant index and scaling proportional to a galaxy's star formation rate (SFR). Unlike previous authors, we do not find any dependence of SNR luminosity on a galaxy's global ISM density. The observed correlation between the luminosity of a galaxy's brightest SNR and a galaxy's SFR can be completely explained by statistical effects, wherein galaxies with higher SFR more thoroughly sample the high-luminosity end of the SNR LF. The LF is well fit by a model of SNR synchrotron emission which includes diffusive shock acceleration and magnetic field amplification, if we assume that all remnants are undergoing adiabatic expansion, the densities of star-forming regions are similar across galaxies, and the efficiency of CR production is constant.
\end{abstract}

\section{Introduction}

Surveys for supernova remnants (SNRs) in extragalactic systems have been ongoing for the past three decades, and we are now at the point where we can study SNR populations and use them to extract valuable insight into SNR evolution and the interstellar medium (ISM). There are distinct benefits to extragalactic samples over studies in the Milky Way. In other galaxies, all SNRs will be at approximately the same distance, and therefore do not suffer the distance uncertainties that plague observations of SNRs in our Galaxy. Also, comparing SNRs between galaxies provides a much larger dynamic range in ISM conditions and a longer baseline for understanding how SNR characteristics and evolution depend on ISM density, star formation rate (SFR), etc. As we stand poised for unprecedented depth in radio continuum imaging with the advent of the Expanded VLA and other Square Kilometer Array precursors, it is timely to look back on the samples of extragalactic SNRs in the literature, synthesize them, and attempt to draw physical conclusions so we might know where to go with the next generation of data.

Extragalactic SNRs have been detected at many wavelengths, but most commonly they are selected by their optical or radio emission. Optical surveys generally use narrow-band imaging of the [\ion{S}{2}] and H$\alpha$ emission lines to detect SNRs and distinguish them from \ion{H}{2} regions. The first search of this kind was carried out by \cite{Mathewson_Clarke73} in the LMC, and these techniques have since been used to establish large (5--100) samples of SNRs in the Local Group \citep{Gordon_etal98, Braun_Walterbos93} and beyond \citep[e.g.,][]{Matonick_Fesen97, Matonick_etal97, Blair_Long97, Blair_Long04}. Radio surveys use data at at least two frequencies to measure discrete sources' spectral indices and separate thermal \ion{H}{2} regions from synchrotron-emitting SNRs. These surveys have been successfully carried out with techniques ranging from single-dish observations in the Magellanic Clouds \citep{Filipovic_etal98} to very long baseline interferometry in compact starburst galaxies \citep[e.g.,][]{Lonsdale_etal06}. Our work here makes use of many radio SNR surveys in the literature; see Section~\ref{Sec_samples} for more details. Multi-wavelength studies show that there is limited overlap between optical- and radio-selected samples of SNRs, because remnants will glow more brightly in the radio if they are expanding into dense ISM, whereas optical SNRs are more easily detected in less dense regions where there is less confusion from ongoing star formation \citep{Pannuti_etal00}.

In this paper, we focus on studies of radio SNRs. The radio luminosity of SNRs is due to synchrotron emission coming from cosmic rays (CRs) that have been accelerated by the SNR through first-order Fermi acceleration \citep{Bell78} and are interacting with the SNR's magnetic field. Due to its power-law spectrum, synchrotron emission is brighter and also suffers less contamination from thermal bremsstrahlung emission at lower frequencies; therefore, radio SNRs are typically selected by their 20 cm emission. The 1.45 GHz spectral luminosities of SNRs detected outside our Milky Way currently span five orders of magnitude, from $\sim$10$^{23}$ erg s$^{-1}$ Hz$^{-1}$ in the SMC to $\sim$10$^{28}$  erg s$^{-1}$ Hz$^{-1}$ in Arp 220 (For comparison, the most luminous Galactic SNR Cas A would have a spectral luminosity of 2.8 $\times$ 10$^{25}$  erg s$^{-1}$ Hz$^{-1}$, assuming a 1 GHz flux density of 2723 Jy and a spectral index of $\alpha$ = $-$0.77 as measured by \cite{Baars_etal77} and a distance of 3.4 kpc from \cite{Reed_etal95}). 

In a simple cartoon featuring a spherical explosion and homogeneous material surrounding it, a SNR starts out in a free expansion phase where its shock speed is $\sim$10,000 km s$^{-1}$ and unaffected by the surrounding medium. This phase ends when the SNR has swept up a mass of ISM or circumstellar material that is approximately equivalent to the mass initially ejected by the explosion, implying a typical duration of 100--1000 years \citep[][hereafter BV04]{Berezhko_Volk04}. Due to the short duration of this phase, statistically very few SNRs will inhabit it at any one time. The end of free-expansion is called the Sedov time, and is important to studies of radio SNRs because it marks a rapid period of very efficient particle acceleration (BV04). At the Sedov time, a SNR has a diameter of $\sim$2-20 pc \citep{Gordon_etal99}, depending on the SN ejected mass and density of the ISM that the remnant is expanding into.

The SNR then enters its adiabatic phase where its evolution can be described by the similarity solution of \cite{Sedov59}. The shock slows down during this time as $v_{s} \propto t^{-3/5}$. BV04 show that a SNR's radio luminosity peaks at the Sedov time and then declines throughout the Sedov phase. Most radio SNRs are thought to be observed at approximately the Sedov time or in the Sedov phase \citep[][BV04]{Berkhuijsen86, Gordon_etal98, Gordon_etal99}. The Sedov phase lasts much longer than the free-expansion phase, and therefore it is statistically likely that SNRs will be observed during their adiabatic evolution.

When the shock wave has slowed down to $v_{s} \approx$  200 km s$^{-1}$, the SNR enters a temperature regime where cooling become dynamically important. This transition from the adiabatic phase to the radiative phase takes place at a SNR age of $\sim$3--5$\times$10$^{4}$ and a diameter of 20--50 pc \citep{Blondin_etal98, Woltjer72}. By this time, SNRs will have quite low luminosities in the radio and will be difficult to detect, although they should start to glow brightly in UV, optical, and IR emission lines \citep[e.g.,][]{Weiler_Sramek88}. As the SNR slows down even more, the cosmic rays which emit radio synchrotron will leak out of the SNR (BV04). Eventually, the shock speed becomes equivalent to turbulent fluctuations in the ISM, and the SNR merges back into the ambient medium. 

Of course, there are many complications that affect an SNR's evolution; not every SNR spends time in each of the four cartoon phases, and oftentimes the transitions between the phases last longer than the phases themselves \citep{Jones_etal98, Reynolds05}. 

Here, we study the populations of radio SNRs in 19 nearby galaxies via their luminosity functions (LFs). The LF can shed light on the evolutionary state of observed SNRs, particle acceleration physics, and magnetic fields in SNRs. Recently, \cite{Thompson_etal09} used the luminosities of extragalactic SNRs to place upper limits on galaxies' magnetic field strengths and test magnetic field amplification in SNRs. Their work does not fully treat the SNR LF, but instead uses the most luminous SNR in a galaxy to constrain the behavior of the SNR population. With our statistical treatment of the LF we will expand upon their work.

In Section 2, we discuss the selection of SNRs and our efforts to make the samples as homogeneous as possible, and in Section 3 we discuss the basic characteristics of the SNRs' parent galaxies. In Section 4, we measure the SNR luminosity function and compare it across galaxies, and then we describe the LFs in terms of physical models of particle acceleration in Section 5. In Section 6, we investigate the possibility of truncation of the LF at the high-luminosity end. Finally, in Section 7 we summarize our results.
\\
\\
\section{SNR Samples} \label{Sec_samples}

\subsection{Selection Criteria}
We have compiled samples of radio SNRs for 18 nearby galaxies from the literature (listed in Table 1). It is important to note that the SNRs presented here were all selected by their radio emission. The details of the SNR searches at 20 cm can be found in Table 2. The data and the selection criteria used by the original authors were largely heterogeneous, so we homogenized the SNR samples as much as possible by requiring a source to have the following traits in order to be considered an SNR:
\begin{itemize}
\item L-band flux density measurement (near 20 cm) at least 3 $\sigma$ above the local noise.
\item Non-thermal spectral index \citep[$\alpha \leq -$0.2; e.g.,][]{Pannuti_etal02}, in order to eliminate \ion{H}{2} regions from our sample.
\item  A counterpart in a narrow-band H$\alpha$ image, in order to distinguish an SNR from a background radio source (any emission from the background source would be redshifted out of the narrow-band filter).
\end{itemize}
Here, we define $\alpha$ as S$_{\nu} \propto \nu^{\alpha}$ where S$_{\nu}$ is the source flux density at frequency $\nu$. In most cases, spectral indices were determined by comparing measurements at 20 cm and 6 cm. For sources where the 20 cm flux density is $>$3$\sigma$ but the 6 cm flux density is not, we place an upper limit at 6 cm of three times the 1$\sigma$ error in the flux density. We use this upper limit on the 6 cm flux density to place an upper limit on the spectral index, and if this allows us to say that the spectral index is less than $-$0.2, the source is considered non-thermal. 

The sources fulfilling these criteria are presented in Tables 3--20, and their selection is discussed in more detail in the next subsection. 20 cm flux densities were originally observed at frequencies ranging from 1.40 to 1.49 GHz (see Table 2), and have all been scaled to 1.45 GHz using the sources' spectral indices. The errors on the flux densities include both measurement errors and uncertainties in the absolute flux calibration. Unless the original authors noted otherwise, we assume calibration uncertainties of 5\% the source flux density \citep[5\% appears to be a typical calibration uncertainty at the VLA\footnote{The National Radio Astronomy Observatory is a facility of the National Science Foundation operated under cooperative agreement by Associated Universities, Inc.}; e.g.,][]{Ulvestad00}. The samples will be further winnowed down in Section 2.4 using a luminosity criterion with the intention of creating complete samples of SNRs. SNRs in complete samples are marked with asterisks in Tables 3--20.

Whenever possible, we strictly applied our selection criteria, even though this forced us to exclude some objects which the original authors considered to be SNRs. For example, \cite{Gordon_etal99} include sources in their list of SNRs in M33 if they have optical counterparts with [\ion{S}{2}]/H$\alpha >$ 0.4. Some of these sources have spectral indices $\alpha > -$0.2, and therefore were rejected from our sample. We do not have [\ion{S}{2}] imaging for all 18 galaxies, and so this optical emission line criterion can not be generalized to our entire sample. For the purpose of this paper, source selection based on [\ion{S}{2}]/H$\alpha$ is not sufficient to be considered an SNR. However, there are a few galaxies which make the standard observations difficult (see the below notes on the Magellanic Clouds, M82, and NGC 253), and our SNR selection criteria were forced to be flexible in these cases. 

Additionally, we excluded recent radio SNe from our samples; four radio SNe in M83, 41.95+57.5 in M82, SN1981K in NGC 4258, SN 1968D in NGC 6946, and SN 1994I in M51 are intentionally removed from the SNR samples.

\subsection{Notes on Indvidual Galaxies}
Below we discuss the SNR samples of each galaxy, in order of increasing distance. 

\emph{LMC: } The SNR sample in the LMC has significantly worse angular (and spatial) resolution than any of the other sample galaxies because it is based on single-dish data acquired with the Parkes telescope. We considered all sources with an ``SNR'' or ``SNR candidate'' designation in \cite{Filipovic_etal98}; see their work for a description of SNR selection criteria. The Parkes study is a multi-frequency survey, but we required sources to have 1.4 GHz flux densities and spectral indices between 20 cm and 6 cm $\leq -$0.2. We calculated uncertainties based on the instructions in \cite{Filipovic_etal95}, which included calibration errors. Positions also came from \cite{Filipovic_etal95}. 

\emph{SMC: } The data for the SMC are significantly higher resolution than for the LMC because the radio maps have ATCA interferometric data added to the Parkes data. We considered sources with ``SNR'' (probably a \emph{bona fide} SNR) or ``snr'' (SNR candidate) designation in \cite{Payne_etal04}. There were a few objects with SNR/snr designation but with $\alpha > -$0.2 which were eliminated from our sample ($\alpha$ values were taken from the Payne \etal catalog, where they use all available flux densities for the calculation). No flux density errors are given in \cite{Payne_etal04}, but \cite{Filipovic_etal02} estimate that the calibration error can be up to 10\% for extended sources. The images also have an r.m.s. noise of 1.8 mJy/beam. We use the diameters as measured off the 2.4 GHz images by \cite{Filipovic_etal05} to calculate the number of beams each SNR subtends, and then use the following equation to calculate conservative calibration + measurement errors for the 20 cm flux densities:
\[ \sigma_{ S_{1.4}} = \sqrt{(0.1 \times S_{1.4})^{2} + ( N_{bm} \times 1.8\ \textrm{mJy/bm})^{2}}. \]
Here, $S_{1.4}$ is the SNR flux density at 1.45 GHz, $N_{bm}$ is the number of beam areas the SNR extends across, and $\sigma_{S_{1.4}}$ is the total error in the 1.45 GHz flux density. 

\emph{IC 10:} We applied our SNR criteria to the sources in Table 1 of \cite{Yang_Skillman93} and came away with three SNR candidates. Additionally, we include the non-thermal superbubble that is the central subject of their paper. Although this source has an unusually large diameter ($\sim$130 pc) and probably results from multiple SN explosions, we include it here because similar sources in other galaxies would not be resolved (see Table 2) and would appear as typical SNRs. We compared the Yang \& Skillman 20 cm sources with an H$\alpha$ image from \cite{GildePaz_etal03}. 

\emph{M33: } We applied the selection criteria described above to the list of radio sources in \cite{Gordon_etal99}. Most sources were checked for H$\alpha$ counterparts by Gordon et al. However, six radio sources fell out of the range of the Gordon \etal H$\alpha$ images, so we checked them against H$\alpha$ images from the UV/visible Sky Gallery \citep{Cheng_etal97}, downloaded from NED\footnote{The NASA/IPAC Extragalactic Database (NED) is operated by the Jet Propulsion Laboratory, California Institute of Technology, under contract with the National Aeronautics and Space Administration.}. We excluded source 102 because it corresponds to the galactic center. 

\emph{NGC 1569, NGC 4214, NGC 2366, and NGC 4449:} We include the SNR candidates from \cite{Chomiuk_Wilcots09}, who used data at three frequencies (20, 6, and 3.6 cm) to constrain spectral indices, and looked for H$\alpha$ counterparts to the radio sources. For sources with no 6 cm flux density measured, Chomiuk \& Wilcots assumed an upper limit that was 3$\sqrt{2}$ times the local noise. To be consistent with the other surveys included here, we recalculate the spectral indices assuming a 3$\sigma$ upper limit, and this tighter constraint qualifies all five ambiguous SNR/\ion{H}{2} region sources as non-thermal; they are included here as SNRs. 

\emph{NGC 300: } We used the list of non-thermal radio sources in \cite{Pannuti_etal00}, which were all checked for H$\alpha$ counterparts. 

\emph{M82: } We used different selection criteria for M82 because, due to the compact and crowded nature of this galaxy, no 20 cm-selected SNR sample exists for it (higher resolutions can be achieved at higher frequencies). In addition, many SNRs in starburst galaxies suffer severe free-free absorption at 20 cm; objects which are easily observable at 6 cm may become undetectable at 20 cm \citep{Tingay04, Lenc_Tingay06}. Therefore, selection by 20 cm data is unsuitable, and instead we use the list of 5- and 15-GHz selected SNRs from \cite{McDonald_etal02}. All sources in their list have spectral indices $\leq -$0.2 with the exceptions of 44.43+61.8 and 45.52+64.7; we do not include these two sources in our sample. We also exclude 41.95+57.5 because it is a rapidly-fading radio SN. This gives a total of 27 SNRs. 

For the 23 of these objects in \cite{Allen_Kronberg98}, we use their measured flux densities to characterize the radio spectra. In addition, we use their models of each source's radio spectrum to determine where the spectrum turns over at low frequency due to free-free absorption. For frequencies higher than this turnover, we fit the flux density measurements with a straight line to determine the spectral index. The error in the slope of the fitted line corresponds to the error in $\alpha$. We also use this line fit to determine the source's flux density at 1.45 GHz, which allows us to correct for extinction in sources with significant free--free absorption by applying the power-law fit from higher frequencies to 1.45 GHz. 

Four of the SNRs are not measured by Allen \& Kronberg, but are studied by \cite{McDonald_etal02}. 20 cm flux densities for these sources are found by fitting a power law between the 5 and 15 GHz data points at 200 mas resolution, and extrapolating this fit to 1.45 GHz. Again, this method should correct for free--free absorption, but gives larger errors because there are only two data points constraining the spectral index, rather than the 4--6 measurements typically provided by Allen \& Kronberg.

We did not require H$\alpha$ counterparts for this galaxy due to the extremely high dust extinction in the center of M82, and the high density of sources implying that background contamination is unlikely to be significant. Positions for sources come from \cite{McDonald_etal02}. 

\emph{M81: } We used the list of radio sources in \cite{Kaufman_etal87} and applied the criteria described above. All sources included in the Kaufman \etal catalog were checked by the original authors to have H$\alpha$ counterparts. 

\emph{NGC 7793:} We used the list of radio SNR candidates in \cite{Pannuti_etal02}. We eliminated one source with $\alpha > -$0.2. Pannuti \etal checked all of their sources to ensure that they have H$\alpha$ counterparts. 

\emph{NGC 253: } For the same reasons as in M82, we cannot select SNRs in NGC 253 by their 20 cm emission. Instead, in the central $\sim$200 pc of NGC 253, we select SNRs using the 6 cm source list in \cite{Ulvestad_Antonucci97}; \cite{Lenc_Tingay06} show that free-free absorption does not significantly affect 5 GHz flux densities. We measure spectral indices for the sources using all available data between 1.3 and 6 cm as presented in Table 13 of Ulvestad \& Antonucci. A source is considered an SNR if its spectral index is non-thermal and its 6 cm flux density measurements is significant to $>$3$\sigma$. We do not include 5.79-39.0 because it is assumed to be the nuclear region of the galaxy by Lenc \& Tingay. 

In addition, \cite{Ulvestad00} performed a search for SNRs at larger galactic radii, but this study is limited in field of view and sensitivity at large radii due to bandwidth smearing. We did not include the ``wide-field sources'' listed by Ulvestad because the survey was very incomplete at these radii. However, we did include the ``compact circumnuclear sources'' if they had $\alpha \leq -$0.2, and these sources appear in Table \ref{tab:n253} as single-integer ID numbers. Both \cite{Ulvestad_Antonucci97} and \cite{Ulvestad00} included calibration errors in their estimates of flux density uncertainties. 

Unfortunately, we did not have access to a suitable narrow-band H$\alpha$ image for NGC 253. However, all of the radio sources are well within the optical disk of NGC 253 where there are high levels of star formation, and it is unlikely these sources are background. \cite{Ulvestad00} estimates that perhaps one of the sources in the 16-square-arcminute survey area may be a background source.  Additionally, like M82, an H$\alpha$ image of NGC 253 would be plagued by dust extinction and lead us to falsely eliminate many SNRs. 

\emph{M83: } We used the radio flux densities from the 1990 observations listed in \cite{Maddox_etal06}, and applied the criteria listed above. The historical SNe 1923A, 1950B, 1957D, and 1983N are detected in the radio by Maddox et al., but none of these sources fulfill our SNR criteria (and even if they did, they would be intentionally excluded from our sample here). We checked for H$\alpha$ counterparts to candidate SNRs using a SINGG image \citep{Meurer_etal06} downloaded from NED. 

\emph{NGC 4736: } We included sources from \cite{Duric_Dittmar88}, who surveyed the circum-nuclear star-forming ring of this galaxy, if they fit our SNR criteria. We found H$\alpha$ counterparts in an image from \cite{Knapen_etal04} downloaded from NED. 

\emph{NGC 6946:} We used the list of discrete radio sources from \cite{Lacey_etal97}, and applied the criteria as described above. SN 1968D corresponds to source 82 in Lacey et al., and is excluded from our sample. We utilized H$\alpha$ images from SINGS \citep{Kennicutt_etal03} accessed through NED. 

\emph{NGC 4258: } We include in our SNR sample the sources in \cite{Hyman_etal01} if they meet the above criteria. We did not include SN 1981K \citep{vanDyk_etal92} even though it appeared in the observations of NGC 4258. 

\emph{M51: } We used the list of discrete radio sources in \cite{Maddox_etal07} and imposed the selection criteria describe above. We used an ACS mosaic of M51 in H$\alpha$ acquired by the Hubble Heritage team to check that sources had an H$\alpha$ counterpart. We excluded source 58 which corresponds to SN 1994I. We note a remarkably bright source (104) which, if it is an SNR, is unusually luminous for a galaxy like M51. It is coincident with H$\alpha$ emission and is therefore included here as an SNR. This source is deserving of follow-up observations to investigate if it is indeed a super-luminous SNR.
\subsection{Arp 220} \label{arp220}

Recent VLBI studies have revealed a rich population of discrete non-thermal radio sources in the ultraluminous infrared galaxy (ULIRG) Arp 220 \citep{Smith_etal98, Lonsdale_etal06}. Despite the relatively large distance to Arp 220 (77 Mpc), these studies achieve high spatial resolution of $\sim$1.5 pc. However, a significant fraction of the radio sources are probably SNe, and not SNRs. When \cite{Lonsdale_etal06} compared two 18 cm images of Arp 220 observed a year apart, they detected four new sources in the second-epoch image. From these observations, they estimate an SN rate of $4\pm2$ yr$^{-1}$. In addition, Arp 220's extreme characteristics make it difficult to compare with other more quiescent nearby galaxies. Its very large SFR give it disproportionate influence when we investigate the composite SNR population of nearby galaxies (as in Figure \ref{difflf_scl}) or find linear fits to the data as a function of SFR (as in Section \ref{lf_scaling}.) For these reasons, we do not include Arp 220 in our main sample of galaxies, but throughout our analysis of the SNR LF, we will compare our results from the 18 ``normal'' galaxies with SNR candidates in Arp 220.

As an attempt at an SNR sample, we include here the non-thermal ``long-lived'' sources (with 18 cm flux densities which did not vary significantly over 11 years) from \cite{Parra_etal07}. One ``ambiguous'' source, W15, is also included because Parra \etal state that it is most likely a long-lived source. The flux densities listed in Table \ref{tab:arp220} were calculated by transforming the 18 cm optically-thin synchrotron flux densities from Table 2 of \cite{Parra_etal07} to 1.45 GHz flux densities using the Parra \etal spectral indices. Completeness was determined by the same technique used for the other galaxies as described in the next subsection. 
\subsection{Defining Completeness} \label{completeness}
\begin{figure*}
\centering
\includegraphics[width=13cm]{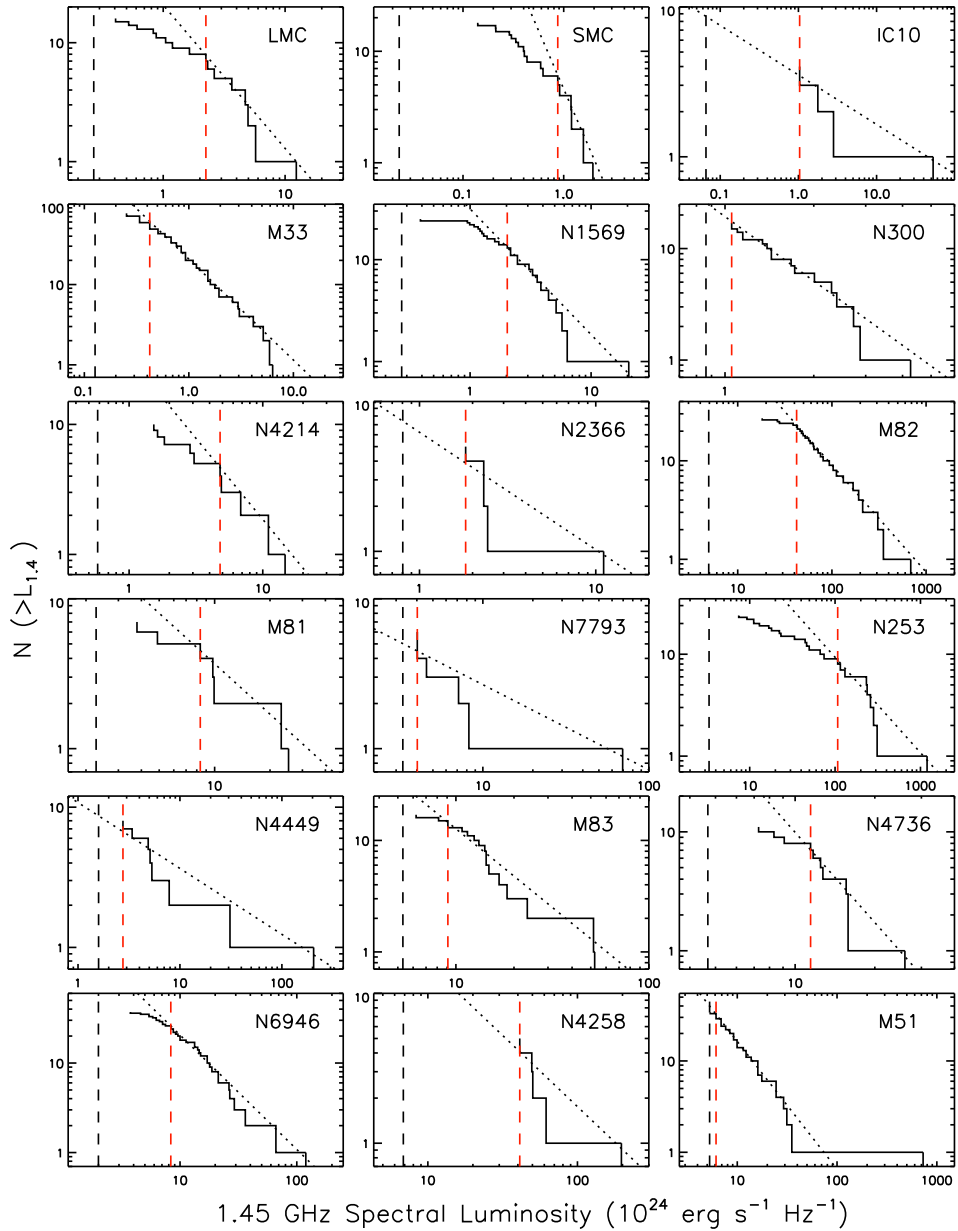}
\caption{Power-law fits (short-dashed lines) to cumulative SNR LFs (solid-line histograms). The true incompleteness limits (where the power laws turn over) are marked with red long-dashed lines, while simple completeness limits defined as 3 times the image noise are denoted by black long-dashed lines.}
\label{fitlf}
\end{figure*}
Completeness can be difficult to define for these samples. Due to the shape of the primary beam, the noise in any interferometric radio image grows as a function of radius from the image phase center. This was an issue for the survey of M33 at 20 cm because of M33's large angular size. At 6 cm, it affects practically every galaxy we present here due to the smaller primary beam at shorter wavelengths. 

In addition, sensitivity may drop off even more quickly with radius due to chromatic aberration; if a galaxy is imaged with wide frequency channels (e.g., ``continuum mode'' with the current correlator at the VLA), sources at significant distances from the image center are smeared radially, spreading their flux over a large area and making them more difficult to detect and measure. This was a significant problem in NGC 253, NGC 4214, and NGC 4258. In these cases, SNR surveys were carried out over regions with less severe bandwidth smearing (typically out to a radius $\approx 2\ \theta_{syn}\ \nu_{0}\ \Delta \nu^{-1}$, where $\theta_{syn}$ is the half-power beam width of the synthesized beam, $\nu_{0}$ is the frequency of the observations, and $\Delta \nu$ is the frequency channel width), and it is these regions that are listed as the survey fields of view in Table \ref{tab:obs}. We recognize that we are not surveying the entire galaxy for SNRs, and correct the relevant galaxy parameters like SFR for this (see Section \ref{galpar}).

SNRs are usually point sources at the distances surveyed here, but occasionally they are resolved. When this is true, the surface brightness of the remnant, not just the total flux, can determine if it is detected. Surface brightness limitations affect nearby high-resolution SNR samples like the SMC and M33 most severely.
 
Additionally, if an SNR is in an environment with vigorous star formation and high backgrounds, it will be significantly more difficult to detect. With interferometric observations, the brightness of the diffuse background emission varies depending on the \emph{uv} coverage of the map. For example, if a galaxy is only observed with the VLA in its most extended configuration, we will only be sensitive to objects with small angular scales, the background will be resolved out, and SNRs should be fairly easy to detect. The maps used in this study were made with a wide range of VLA configurations, and therefore the problem of high backgrounds will vary quite a lot from galaxy to galaxy.

The spatial resolutions of the SNR surveys vary from 13 pc to 221 pc. To investigate the impact of resolution on the SNR catalogs, we produced a catalog of discrete 20 cm sources in NGC 6946 observed with the VLA in its B configuration, and compared it with the \cite{Lacey_etal97} catalog of radio sources detected in the A-configuration. The B-configuration images have approximately three times lower resolution, and more sensitivity to diffuse emission. Of the 37 SNRs found in the higher resolution images, only four were not detected in the B-configuration images because of poor resolution and confusion with other sources. Another four SNRs were not detected because of the higher diffuse background in the B-configuration images. Therefore, it appears that resolution affects SNR samples only mildly. 

The observations of each galaxy in our sample have distinct limitations, and therefore only a subset of the SNRs described in Section 2.2 can be considered to be in complete samples. We define completeness in terms of SNR spectral luminosity using a method developed by \cite{Gordon_etal99} on the M33 SNR LF. We assume that the cumulative LFs of SNRs can be described as single power laws, and departures from them at the faint end are due to incompleteness. For each galaxy, we plot the cumulative SNR LF as measured at 20 cm and fit a power law to it at the bright end. At fainter luminosities, most galaxies' LFs exhibit a break where they can no longer be described by this power law and must be fit with a shallower one. We call this break in the cumulative LF the true completeness limit for the sample (see Appendix A for more details on how the location of this break is determined). 

Our cumulative LFs and power-law fits are shown in Figure \ref{fitlf} for our
18 sample galaxies. The low-luminosity limits where the LFs cease to be fit by
power laws correspond to our completeness limits (red dashed
line). Additionally, Figure \ref{fitlf} shows the flux density limit at which
one might expect to reliably measure point sources, at 3 times the image
noise. For most galaxies,  the completeness limit is significantly brighter
than the survey flux density limit, due to the reasons described above. Our
definition of completeness appears to be conservative, and unfortunately
limits quite drastically the number of SNRs in our sample for some galaxies
(see Tables 3--21, where SNRs in complete samples are marked with
asterisks). In total, we have 259 SNRs in complete samples across the 18
sample galaxies, and four SNRs in the Arp 220 complete sample. 
\section{Parent Galaxy Parameters} \label{galpar}
Some basic characteristics for our 18 sample galaxies are listed in Table 1. All are nearby ($\leq$ 8.0 Mpc) and star-forming. They span a range of Hubble types from Irregular to Sab.

We adopt the distances listed in \cite{Kennicutt_etal08}, which are mostly drawn from \cite{Karachentsev_etal04}. We apply errors on these distances as prescribed by Karachentsev et al.: 5\% uncertainty for distances found using Cepheid variable stars, 10\% for distances derived from the tip of the red giant branch, and 20\% for distances found from supergiant stars. The distance to M51 was measured using surface brightness fluctuations of its companion, and the error on its distance measurement is 13\% \citep{Tonry_etal01}. Absolute B-band magnitudes also come from \cite{Kennicutt_etal08}, using the distances listed in Table 1. These correct for Galactic extinction by using the average of values supplied by \cite{Burstein_Heiles78} and \cite{Schlegel_etal98}. 

We calculate SFRs using a combination of H$\alpha$ and mid-infrared data as directed by Equation 7 in \cite{Calzetti_etal07}:
\begin{equation}
\textrm{SFR} (\textrm{M}_{\odot}\ \textrm{yr}^{-1}) = 5.3 \times 10^{-42}\ \left[ L(H\alpha) + 0.031\ L(\textrm{24$\mu$m}) \right]
\end{equation}
where L(H$\alpha$) and  L(24$\mu$m) are luminosities in erg s$^{-1}$, and the IR luminosity is converted from a spectral luminosity by multiplying by the frequency. This prescription corrects the H$\alpha$ luminosity for extinction using the 24 $\mu$m emission (here we actually use 25 $\mu$m emission from the IRAS satellite). We use the H$\alpha$ luminosities listed in \cite{Kennicutt_etal08}, except for IC 10, whose H$\alpha$ luminosity comes from \cite{Hunter_Elmegreen04}. The majority of 25 $\mu$m flux densities come from \cite{Sanders_etal03}, with a few measured by \cite{Rice_etal88}, \cite{Lisenfeld_etal07}, and \cite{Moshir_etal92}. There are no IRAS measurements for IC 10, so we used an H$\alpha$-only SFR for this galaxy, assuming an internal+foreground reddening of E(B-V) = 0.85, consistent with values in the literature \citep{Hunter01}.

Not all galaxies are completely covered by the SNR surveys used here, and therefore it is not consistent to compare SFRs measured from integrated luminosities with the SNR populations derived from these surveys. In all cases except M33, M82, and NGC 253, we use archival 24 $\mu$m Spitzer images to measure the fraction of the SFR inside the SNR survey area. The disk of M33 was not entirely covered by Spitzer observations, so in this case we used 25 $\mu$m IRAS observations to trace the star formation. In the cases of M82 and NGC 253, archival Spitzer/MIPS images were saturated, and we were forced to use 20 cm radio continuum images from \cite{Condon87} to trace the SFRs. For each galaxy, the fraction of its star-formation activity included in its SNR survey area is listed in Table 2.

Previous studies of extragalactic SNRs have correlated the characteristics of SNRs with the global ISM densities of their parent galaxies \citep{Hunt_Reynolds06, Thompson_etal09}. For comparison with these studies, we calculate rough estimates of the ISM density for each galaxy ($\rho_{0}$) using the SFRs in the survey areas and the Schmidt-Kennicutt law. We could use measurements of the gas surface density ($\Sigma_{g}$) made directly from \ion{H}{1} and CO measurements, but these quantities are usually measured over the entire galaxy, and it would be very difficult to calculate them in the SNR survey areas (this is particularly important in cases like NGC 4736, where the galaxy is only gas-rich and star-forming in a rather small area, and calculating the ISM density over the entire disk would dramatically dilute the true ISM density in the region producing SNRs). Therefore, we instead convert SFRs into SFR surface densities ($\Sigma_{SFR}$), and then use the relationship from \cite{Kennicutt98} to find $\Sigma_{g}$:
\begin{equation}
\Sigma_{g} = 374.1\ \Sigma_{SFR}^{0.71}
\end{equation}
We assume that all galaxies have a constant gas scale height of 100$\pm$50 pc \citep{Thompson_etal09} although this assumption probably contributes systematic errors, especially given evidence that the gas scale height is a function of Hubble type \citep{vandenBergh88, Brinks_etal02}. We then use the inclinations in Table 1 to find path lengths through the galaxies and divide $\Sigma_{g}$ by the path length to find an ``average'' volume density; these values of $\rho_{0}$ can be found in Table 1. We recognize that this method for calculating density is very rough, and should only be interpreted as an approximate diagnostic.

Basic data on Arp 220 are also listed in Table 1. We use a SFR of 127 M$_{\sun}$ yr$^{-1}$ (as found by \cite{Anantharamaiah_etal00}, but converted to the IMF used by \cite{Calzetti_etal07}). We assume that 100\% of the star formation activity is taking place in the SNR survey area, which is probably a slight overestimate. 
\section{Radio SNR Luminosity Functions} \label{radiolf}
\begin{figure*}
\centering
\includegraphics[width=10.5cm, angle=90]{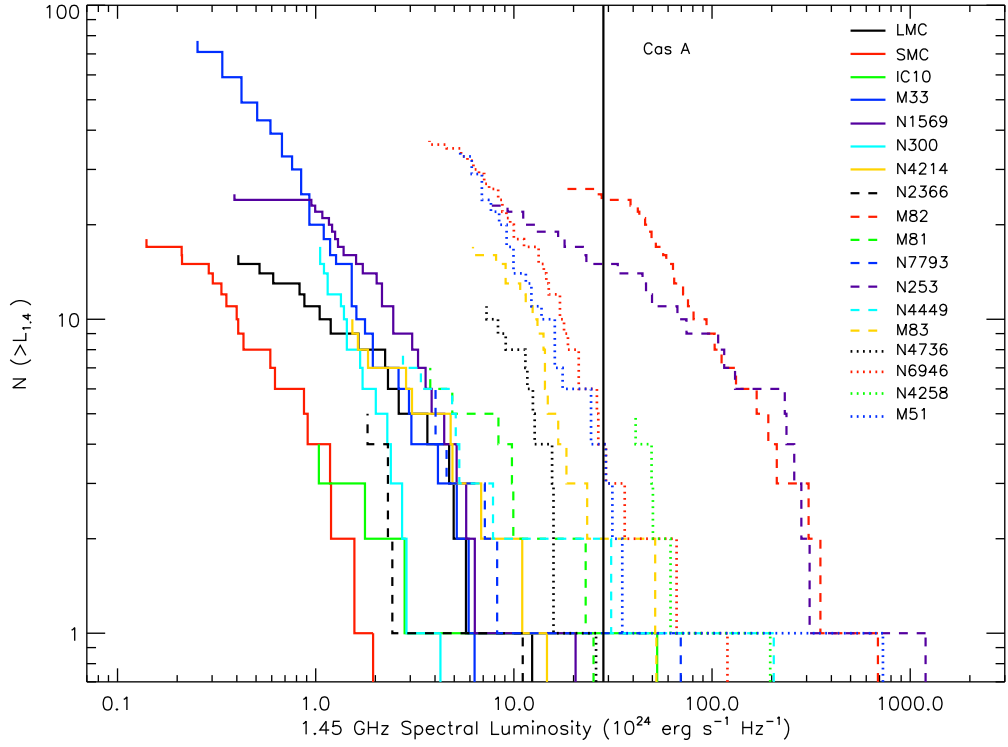}
\caption{Cumulative 20 cm luminosity functions for SNRs in the 18 sample galaxies. The spectral luminosity of Galactic SNR Cas A is also marked as a solid vertical line.}
\label{cumlf}
\end{figure*}
\begin{figure*}[htp]
\centering
\includegraphics[width=10.5cm, angle=90]{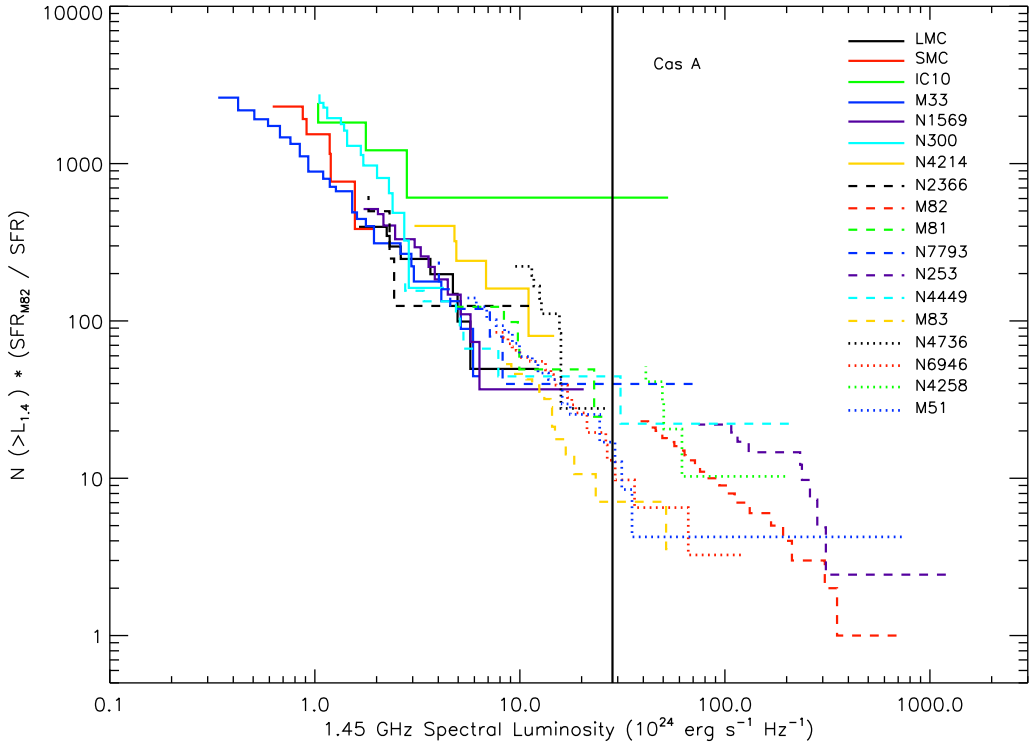}
\caption{Cumulative SNR 20 cm luminosity functions scaled by the inverse of each galaxy's SFR and then normalized by the SFR of M82. Only SNRs in the complete samples are included. The spectral luminosity of Galactic SNR Cas A is also marked.}
\label{cumlf_scl}
\end{figure*}
To convert our SNR flux densities to spectral luminosities, we assume the distances to the galaxies in Table 1, and use the equation:
\begin{equation}
L_{1.4} = 1.20 \times S_{1.4} \times D^{2}
\end{equation}
where D is the distance in Mpc, S$_{1.4}$ is the 1.45 GHz flux density of an SNR in mJy, and L$_{1.4}$ is the SNR spectral luminosity in units of 10$^{24}$ erg s$^{-1}$ Hz$^{-1}$. Errors in the spectral luminosities are calculated as:
\begin{equation}
\sigma_{L_{1.4}} = \sqrt{(1.20 \ D^{2} \ \sigma_{S_{1.4}} )^{2} + (2.40 \ S_{1.4} \ D \ \sigma_{D})^{2}}
\end{equation}
where $\sigma_{D}$ is the error in the distance as described in the previous section, and $\sigma_{S_{1.4}}$ are the errors on the flux density measurements as listed in Tables \ref{tab:lmc}--\ref{tab:arp220}. 
 
Figure \ref{cumlf} shows cumulative luminosity functions for the SNRs in our 18 sample galaxies. The SNR spectral luminosities span almost four orders of magnitude, from $\sim$10$^{23}$ erg s$^{-1}$ Hz$^{-1}$ in the SMC to almost 10$^{27}$  erg s$^{-1}$ Hz$^{-1}$ in M82 and NGC 253. Unfortunately many of the galaxies' SNR samples do not overlap in luminosity space. For example, the most luminous SNR in M33 is fainter than the least luminous SNR observed in M82. The galaxies which host the more luminous SNRs tend to be crowded and/or at relatively large distances, meaning that the SNR surveys of these galaxies will become incomplete at higher luminosities. Therefore it can be difficult to directly compare SNR populations between galaxies. A similar problem was presented for high-mass X-ray binaries in nearby galaxies by \cite{Grimm_etal03}, and much of our analysis is inspired by theirs.

The galaxies with the highest SFRs host the most luminous SNRs; in M82 and NGC 253, many of the SNRs are more luminous than Cas A. Can the differences between the luminosities of SNR populations be completely explained by differences in SFR? Galaxies with higher SFRs will host SNe explosions more often, and therefore the luminous end of the SNR LF will be more thoroughly populated. We also expect the total number of SNRs in a galaxy to scale with the SFR. In Figure \ref{cumlf_scl} we again plot the cumulative LFs, but we scale each LF by the inverse of its parent galaxy's SFR. Most SNRs adhere to a straight line in this diagram, implying that scaling by SFR removes most of the differences between LFs. One notable exception is the most luminous object in IC 10. It sits significantly above the line defined by the rest of the SNRs, implying that it is a much more luminous object than would be expected for a galaxy with the SFR of IC 10. This is not surprising, as \cite{Yang_Skillman93} identified the source as a superbubble and state that it is probably powered by multiple SNe explosions. Because this object is so clearly aberrant, and because there is a recognized reason for its outlier status, we exclude it from further analysis. We also note that the high-luminosity end of this plot shows larger scatter than the low-luminosity end, implying that the SNR LF is not well sampled at high luminosities in many galaxies.
\subsection{LF Power-Law Index ($\beta$)} \label{lf_index}
We can constrain the shapes of the differential SNR LFs if we make the simple assumption that all LFs can be fit with a single power law. We write such an LF as:
\begin{equation}
n(L_{1.4}) = {{dN}\over{dL_{1.4}}} = A \ L_{1.4}^{\beta}
\end{equation}
where $n(L_{1.4})$ is the number of SNRs with spectral luminosity $L_{1.4}$, $A$ is a scaling constant, and $\beta$ is a negative number. We determine the power-law index using the maximum likelihood estimator as described in \cite{Clauset_etal07}:
\begin{equation}
\hat{\beta}^{\prime} = -1 - n\left[\sum_{i=1}^n \textrm{ln}{{L_{1.4,i}}\over{L_{1.4,min}}}\right]^{-1}
\end{equation}
For each galaxy, $n$ is the number of SNRs in the complete sample, and $L_{1.4,min}$ is the spectral luminosity of the least-luminous SNR in the complete sample. However, for small sample sizes, $\hat{\beta}^{\prime}$  will be biased low (more negative) compared to the true $\beta$ value. We can correct for this bias with a simple analytic expression:
\begin{equation}
\hat{\beta} = \left({{n-1}\over{n}}\right)\ \left(\hat{\beta}^{\prime} - {{1}\over{n-1}}\right)
\end{equation}
The standard error on $\hat{\beta}$ is also prescribed by Clauset \etal:
\begin{equation}
\sigma_{\hat{\beta}} = (-\hat{\beta} - 1) \ {{n}\over{(n-1) \ \sqrt{n-2}}}
\end{equation}
\begin{figure}
\centering
\includegraphics[width=7cm, angle=90]{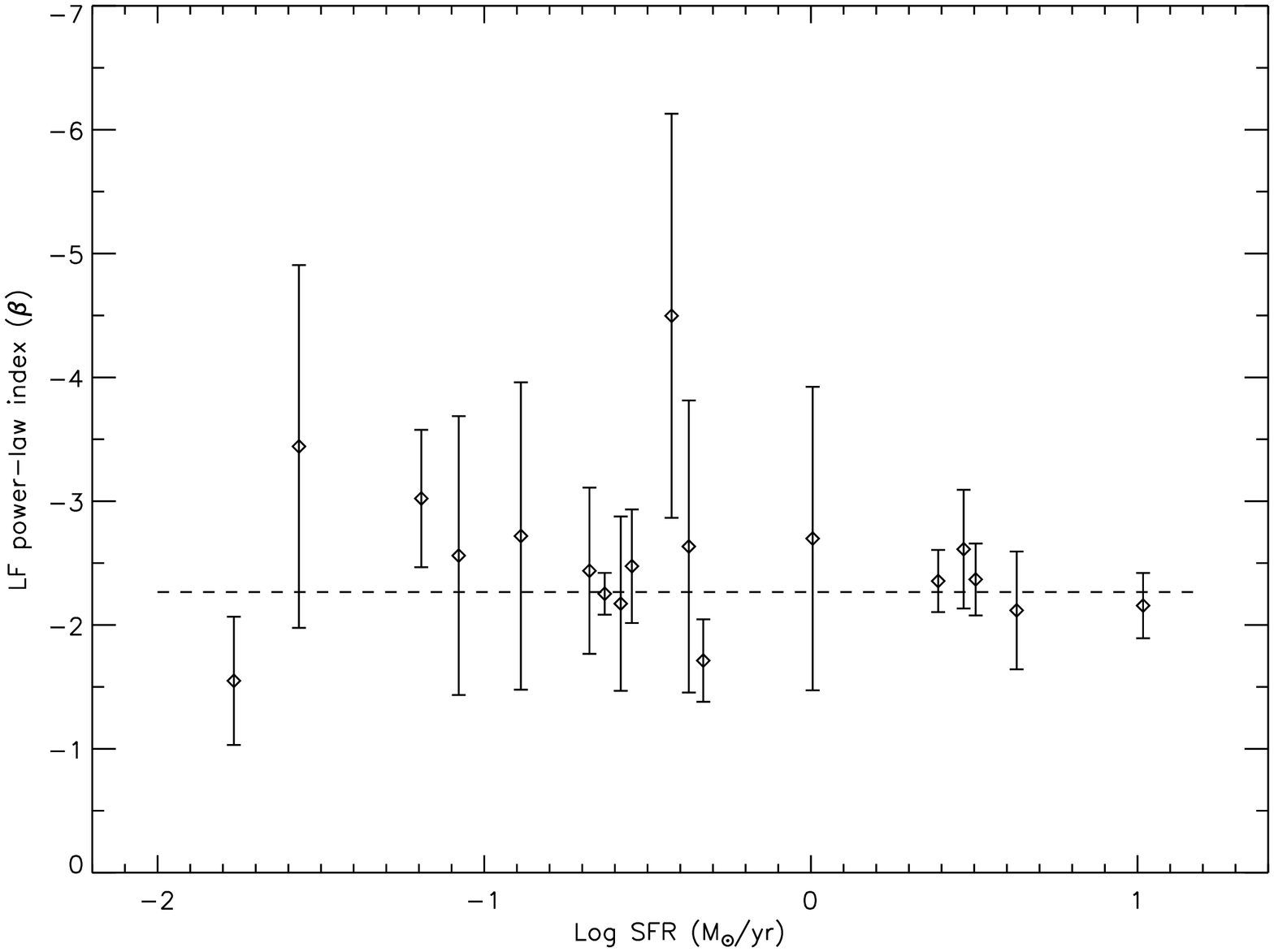}
\caption{Best-fit values of $\beta$ (the power-law index) plotted against SFR for each galaxy. Error bars represent 1$\sigma$ uncertainties in $\beta$ determinations. The weighted-average $\beta$ is marked with a dashed line.}
\label{pind}
\end{figure}

Values of $\hat{\beta}$ range from $-$1.5 to $-$4.5 between our 18 sample galaxies, but many of the values have large uncertainties due to small SNR sample sizes. Determinations of $\hat{\beta}$ can be found in Table \ref{tab:lfs}. Figure \ref{pind} plots the power law index for each galaxy against the galaxy's SFR. The $\hat{\beta}$ values are all consistent with one another, and there is no evidence for systematic trends in $\hat{\beta}$ with SFR (A Spearman rank correlation test gives a correlation coefficient of r$_{s}$ = 0.29 or a two-tailed $p$-value of 0.25). The weighted-mean value for the power-law index is $\bar{\beta} = -2.26 \pm 0.10$. As can be seen in Figure \ref{pind}, almost all of the galaxies fall within $\lesssim$1$\sigma$ of the mean, and no $\hat{\beta}$ deviates by more than 1.7$\sigma$. The galaxies with the largest sample sizes (N $>$ 20) and the best defined SNR LFs--- M33, M82, NGC 6946, and M51--- all have very similar LF indices in the range $\hat{\beta}$ = $-$2.1 to $-$2.4. The SNRs from Arp 220 could also be drawn from the same power law, as $\hat{\beta} = -3.00\pm1.89$ for this galaxy. All of the current data are consistent with being drawn from a single power law. 

\begin{figure*}
\centering
\includegraphics[width=12cm, angle=90]{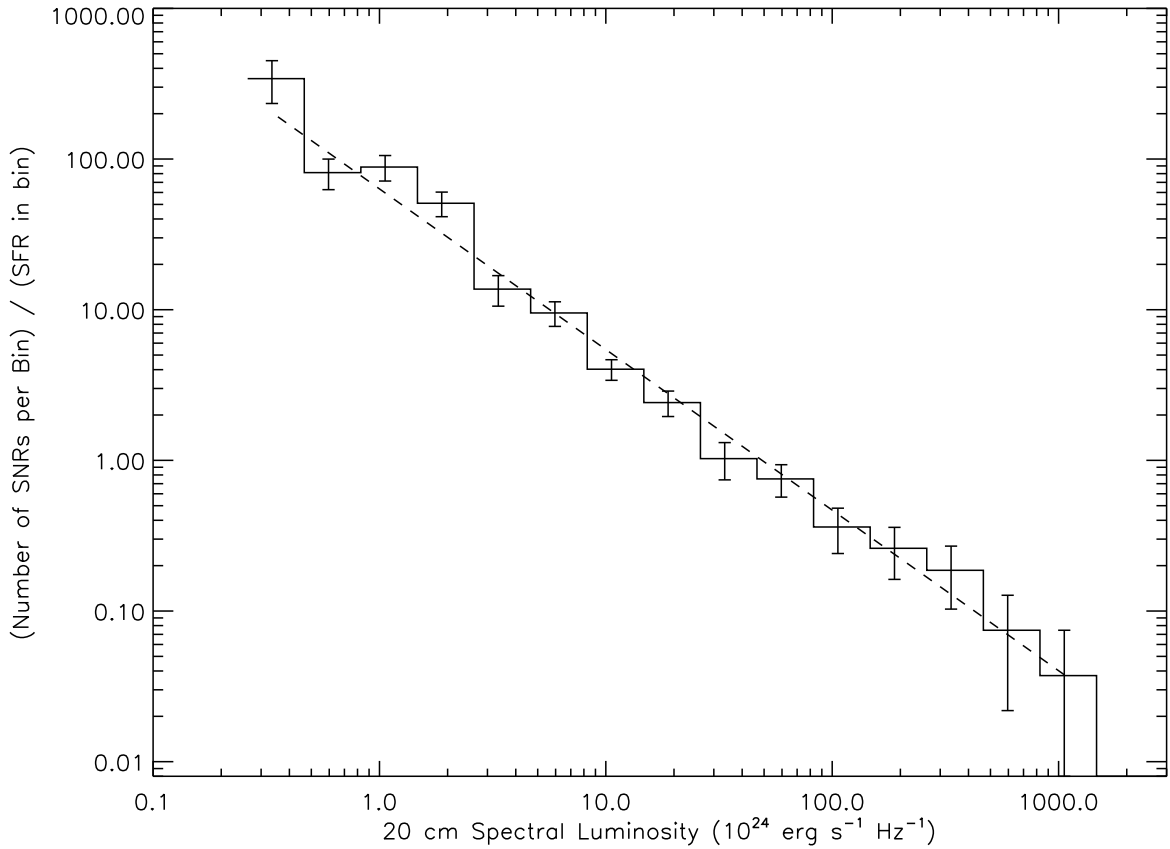}
\caption{Composite SNR LF including the SNR samples from all 18 sample galaxies. To correct for variable completeness between bins, the number of SNRs in each bin is scaled by the totaled SFR for all galaxies which are complete in that bin. Error bars represent simple Poissonian uncertainties. The dashed line is the best power law fit with $\beta=-2.07$.}
\label{difflf_scl}
\end{figure*}

Additional evidence can be found for this assertion if we combine data from
the 18 sample galaxies to make a composite SNR LF, as can be seen in Figure
\ref{difflf_scl}. Because all of the surveys are sensitive to SNRs in the
brightest bins, but only a few surveys are sensitive to SNRs in the least
luminous bins, we had to scale each bin to correct for variable
completeness. For a given bin, we summed up the SFRs for all galaxies whose
SNR surveys are complete in that bin, and then we scaled the number counts in
the bin by the inverse of this sum. This produces a smooth power law over
almost four orders of magnitude in spectral luminosity. The power law is best
fit with an index $\beta = -2.07 \pm 0.07$ (where the uncertainty is found
using the true number of SNRs, $n$ = 258). This is consistent within the
uncertainties with $\bar{\beta}$ found by fitting each galaxy individually and
taking the weighted mean.
\subsection{LF Power-Law Scaling ($A$)} \label{lf_scaling}
\begin{figure*}
\centering
\includegraphics[width=13cm]{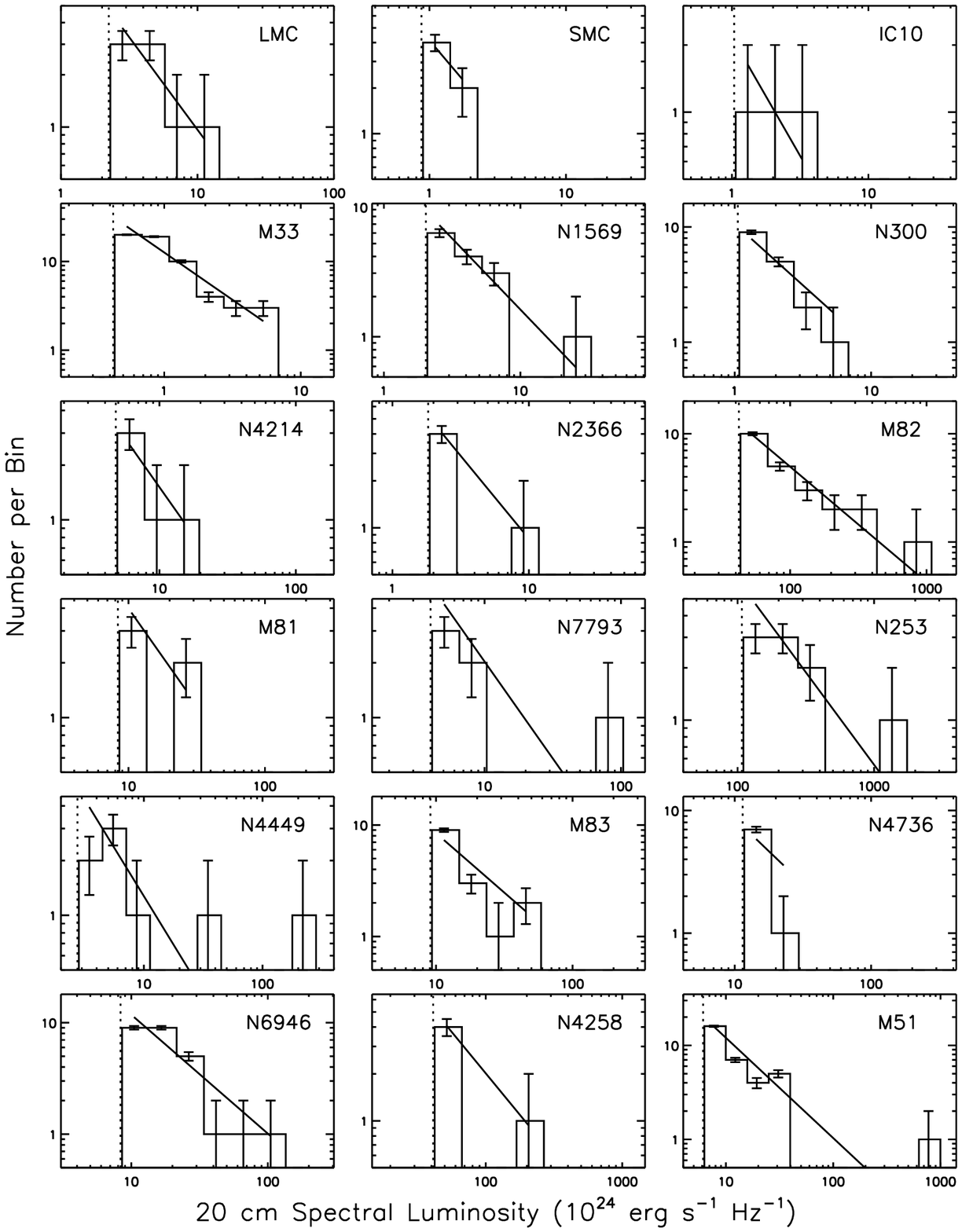}
\caption{Differential SNR LFs for each of the 18 sample galaxies and power-law fits drawn with solid lines. All LFs are shown over two orders of magnitude in spectral luminosity, except for NGC 4449 and M51 which have unusually large dynamic ranges. Power law indices are held fixed at $\beta=-2.07$ and the scaling is allowed to vary. Vertical dotted lines represent the true completeness limits of the SNR samples.}
\label{difflf}
\end{figure*}
As no maximum likelihood estimator is calculable for $A$, the normalization constant is found by binning the data into spectral luminosity bins, and calculating the scaling that is needed to make $L_{1.4}^{\beta}$ match the observed number of sources in each bin. We impose the power-law index determined from the composite LF, $\beta = -2.07$, upon all of the galaxies. To determine $A$, we then take the weighted average of the values found for each bin. To constrain the uncertainty on $A$, we run Monte Carlo simulations by randomly sampling the power law distribution function with the same number of points as are in the complete SNR samples and determining $A$ for these simulated data sets. After 10$^{4}$ runs per galaxy, we can determine the uncertainty on $A$ expected only due to small-number statistics. We then add this in quadrature with the standard error of the weighted mean determined from the data. Values for $A$ are listed in Table \ref{tab:lfs}. In NGC 4449, we excluded the brightest SNR from our calculation of $A$ because it is a severe outlier in a small sample, and therefore disproportionately affects $A$.
 
We plot the differential LFs for our 18 sample galaxies in Figure \ref{difflf}
and overplot the LF fits assuming $\beta = -2.07$ and $A$ as described
above. It is clear that many of the SNR samples suffer from small number
statistics, but by and large the LFs are well-described by the power-law
formulation applied here. 
\begin{figure}[htp]
\centering
\includegraphics[width=9cm]{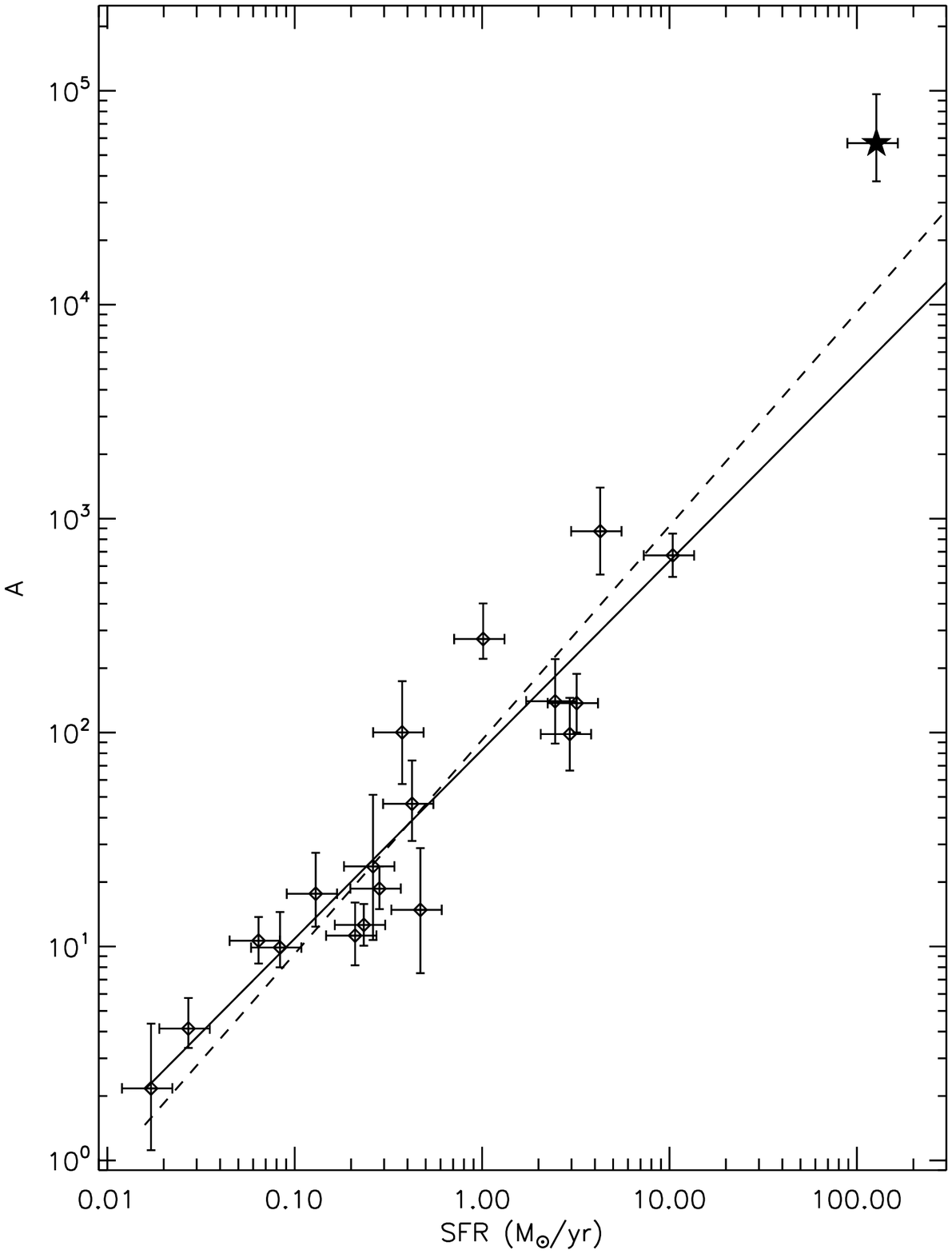}
\caption{The LF scaling factor $A$ as a function of SFR. The solid line is the best fit to the data (slope = 0.88), while the dotted line has a slope of 1.0, representing a linear scaling of $A$ with SFR. Arp 220 is marked with a star and is excluded from the line fit. 30\% uncertainties are plotted as error bars on SFR values.}
\label{pscl}
\end{figure}
 
First and foremost, we would expect the 20 cm luminosity of a SNR population to depend on a galaxy's SFR if we assume that most radio SNRs are from the core collapse of massive stars. In Figure \ref{pscl}, we plot $A$ against galaxy SFR, and see that it correlates well with SFR. We fit log $A$ as a function of log SFR by bootstrapping linear least-squares fits to the data, excluding Arp 220 from the fit because of the reasons discussed in Section \ref{arp220}. We find:
\begin{equation}
A = \left(83^{+18}_{-15}\right) SFR^{0.88\pm0.08},
\end{equation}
which is marked by the solid line in Figure \ref{pscl}. A simple model for the SNR LF predicts that A should be linearly proportional to the SFR (see Section 5.2), and our fit to the data is consistent with this hypothesis. If we impose linear proportionality between A and SFR, then the best fit to the data is marked by a dashed line in Figure \ref{pscl} and is expressed as:
\begin{equation}
A = \left(92^{+17}_{-13}\right) SFR.
\end{equation}

Therefore, the current data are consistent with a power-law SNR LF with constant power-law index across galaxies and scaling that is proportional to SFR. More attention will be given to the astrophysical implications of the LF in the next section. We also note that Arp 220 (which was excluded from the fit) appears to be an outlier, with unusually high $A$ for its SFR, and this too will be discussed in the next section.
\section{Physical Models for the SNR LF}

The synchrotron emission from radio SNRs is dependent on both the cosmic ray energy and the magnetic field energy in the remnant as described by \cite{Longair81}:
\begin{equation}\label{eqSNRlum}
L_{\nu} \propto a(\alpha)\ V\ K\ B^{-\alpha+1}\ \nu^{\alpha}
\end{equation}
where $\alpha$ is the spectral index of the synchrotron emission as defined in Section 2.1; $a$ is a constant that is dependent on $\alpha$ and can be found in Table 18.1 of \cite{Longair81}; $B$ is the magnetic field strength in the SNR; $V$ is the volume of the SNR; and $K$ is the scaling factor of the CR electron energy distribution, defined as $N(E)\ dE = K\ E^{2\alpha - 1}\ dE$, where N(E) is the number density of CR electrons of a certain energy in the remnant. Here, we will assume the CR electron energy spectrum can be described as a power law E$^{-2}$ which gives a synchrotron spectral index of $\alpha = -0.5$.

In many systems, the minimum energy assumption is used to tease out the relative contributions of CRs and magnetic fields to the synchrotron emission. However, there is really no physical reason for assuming equipartition in SNRs \citep{Jones_etal98}, and it is likely that the energy in the magnetic field is only a few percent of the energy in relativistic particles \citep{Hillas05}. We therefore do not make the minimum energy assumption here.

BV04 model CR production and magnetic field amplfication in SNRs with a full non-linear treatment of diffusive shock acceleration. They find that CR production peaks dramatically at the end of an SNR's free expansion phase, and during the Sedov phase the energy in CRs is approximately constant. Adiabatic losses to the CR energy are presumably countered by low-level ongoing CR acceleration. The relatively low-energy CR electrons which emit synchrotron in the radio (typical energies of $\sim$ 1--10 GeV) are contained by the SNR until the shock wave slows to less than the speed of CR diffusive propagation. This roughly corresponds to the radiative snowplow phase, at an SNR age of $\sim$10$^{5}$ years \citep{Hillas05}. We also note that synchrotron losses are unlikely to be important at 20 cm given the predictions of BV04 that the amplified field strength is 10--100 $\mu$G in the Sedov phase. \cite{Thompson_etal06} state that the synchrotron cooling timescale is:
\begin{equation}
\tau_{syn} = 8.3 \times 10^{5}\ B_{100}^{-3/2}\ \textrm{yr}
\end{equation}
at 1.45 GHz, where $B_{100}$ = $B$/100$\mu$G. This is significantly longer than the duration of the Sedov phase (a few $\times$ 10$^{4}$ years).

The CR energy content of an SNR is only weakly dependent on the ambient ISM density according to BV04. In their models, increasing the ISM density by three orders of magnitude only increases the CR energy by a factor of two in  the Sedov phase (from 20\% to 60\% of the SN energy). This can also be seen if we evaluate the CR energy using the simple test-particle assumption of \citet[Equation 10]{Bell78} at the Sedov time (as the Sedov time is when the vast majority of CRs are produced). CR energy density is proportional to ISM density, but the SNR volume at the Sedov time $\propto \rho_{0}^{-1}$; the two factors cancel, and the total CR energy is independent of density.

Therefore, if we assume that most of the radio SNRs imaged in external galaxies are in their Sedov phase, we can assume that their CR energy is roughly independent of time and ISM density. CR energy is simply a fraction of the SN explosion energy ($E_{SN}$). If we in turn assume that $E_{SN}$ is roughly constant, then the synchrotron emission only depends on the magnetic field strength.

\subsection{Magnetic Field Compression Scenario}
If, in a typical radio SNR in its Sedov phase, the magnetic field is not amplified in the SNR but is instead simply compressed by the shock wave, then a compression factor ($f$) describes the magnetic field in the SNR as a multiple of the magnetic field strength in the ambient ISM ($B_{0}$). In the case of a strong shock passing through a randomly oriented magnetic field, $f$ = 3.32 \citep{Reynolds_Chevalier81}. When non-linear effects are taken into account, the shock may become significantly modified and the compression factor may reach $f \approx$ 6 \citep[BV04]{Volk_etal02}. Under these assumptions, the magnetic field strength in an SNR is approximately constant throughout the Sedov phase, at most varying by a factor of two. When combined with the constant CR energy, this implies that the radio spectral luminosity should not vary throughout the Sedov phase \citep{Reynolds_Chevalier81}. The most significant parameter determining a SNR's synchrotron luminosity is the magnetic field strength in the surrounding ISM. 

We can rewrite the expression for the SNR LF as:
\begin{equation}
{{dN}\over{dL_{1.4}}} = {{dN}\over{dB_{0}}}\ \left({{dL_{1.4}}\over{dB_{0}}}\right)^{-1}.
\end{equation}
${{dN}\over{dB_{0}}}$ describes the probability density (actually, the number) of SNe exploding into an ISM with a given magnetic field strength. Let us assume a power law form for it:
\begin{equation}
{{dN}\over{dB_{0}}} = D\ B_{0}^{\eta}.
\end{equation}
Using  equation \ref{eqSNRlum}, we assume $V\ K \propto E_{SN}$ as described above, and  write ${{dL_{1.4}}\over{dB_{0}}}$ as:
\begin{equation}
{{dL_{1.4}}\over{dB_{0}}} \propto E_{SN}\ f^{1.5}\ B_{0}^{0.5}.
\end{equation}
And finally, from equation \ref{eqSNRlum}, we know that $B_{0} \propto E_{SN}^{-2/3}\ f^{-1}\ L_{1.4}^{2/3}$, so we find:
\begin{equation}
{{dN}\over{dL_{1.4}}} \propto E_{SN}^{(-\eta-1)/1.5}\ f^{(-\eta-1)}\ L_{1.4}^{(\eta-0.5)/1.5}.
\end{equation}
Therefore, $\beta = (\eta-0.5)/1.5$. Assuming that $\beta$ = $-$2.07 as we found in Section 4, this implies that $\eta$ = $-$2.6. The SNR LF could be used to constrain the distribution of magnetic field strengths in star-forming regions. It should be noted that this interpretation of the SNR LF implies a large, although not entirely unrealistic, dispersion in ISM magnetic field strengths. The SNRs in M33 range over a factor of 25 in luminosity, implying a dispersion in ambient magnetic field strength of $\sim$9 (if we hold E$_{SN}$ and $f$ constant). Of course, if there are many low-luminosity SNRs in M33 which are not currently observable, the implied dispersion in magnetic field strength could be significantly higher. The luminosities of SNRs in the 18 sample galaxies vary by almost four orders of magnitude, translating to a factor of a few hundred in magnetic field strength. 

However, a simple $B$-field compression scenario is probably not realistic. In young SNRs it is well established that the magnetic field must be significantly amplified over the ambient ISM value to fit observations of X-ray synchrotron emission \citep{Volk_etal05}. In the Sedov phase the X-ray synchrotron emission plummets quickly (BV04), and therefore there is no direct test for magnetic field amplification in more evolved SNRs. \cite{Thompson_etal09} used extragalactic radio SNRs as a test of magnetic field amplification by assuming that 1\% of the SN energy goes into CR electrons, and then using the 20 cm luminosities of SNRs to measure the magnetic field strength in SNRs. In normal star-forming galaxies like the ones we study here, they find that the magnetic field strength in SNRs is greater (by factors of a few to 10) than the strongest field obtainable by simple compression of the ISM. They therefore claim that modest $B$ field amplification is taking place in SNRs. These findings only become stronger if we use the estimates of BV04 which imply that only 0.2--0.6\% of SN energy goes into accelerating CR electrons. In this case, the SNR magnetic fields will be stronger by a factor of 1.4--3 than those estimated by Thompson \etal Therefore, magnetic field amplification is likely shaping the SNR LF, and we use it below to develop a physical interpretation of the SNR LF.

\subsection{Magnetic Field Amplification Scenario}
BV04 assume that the magnetic field energy density in an SNR is amplified to a fraction (1\%) of the SNR pressure via the mechanism of \cite{Luceck_Bell00}:
\begin{equation}
B^{2}/(8\pi) = 0.01\ \rho_{0}\ v_{s}^{2}.
\end{equation}
This implies that the magnetic field is weakening as the SNR expands, and therefore the synchrotron luminosity decreases throughout the Sedov phase. We also assume that the energy in CR electrons is a constant fraction of E$_{SN}$ and that $v_{s}$ can be described by the standard Sedov similarity solution $v_{s} \propto (E_{SN}/\rho_{0})^{1/5}\ t^{-3/5}$. Then the spectral synchrotron luminosity scales as:
\begin{equation}\label{eqLtoT}
L_{\nu} \propto E_{SN}^{1.3}\ \rho_{0}^{0.45}\ t^{-0.9}.
\end{equation}
This is consistent with the findings of BV04 despite their more detailed non-linear treatment of particle acceleration. Note that at a given SNR diameter, all SNRs should have roughly the same spectral luminosity, with some spread due to $E_{SN}$. However, remnants in denser media reach the Sedov time (corresponding to their peak luminosity) when they still have relatively small diameters. Therefore, their peak spectral luminosity is brighter than that of SNRs in lower density media, and they continue to be more luminous through much of the Sedov phase (see Figure 4 of BV04). 

In this case, we can write the SNR luminosity function as:
\begin{equation}\label{eqLFdiff}
{{dN}\over{dL_{1.4}}} = {{dN}\over{dt}}\ \left({{dL_{1.4}}\over{dt}}\right)^{-1}.
\end{equation}
${{dN}\over{dt}}$ is the production rate of SNRs; if we assume that most SNRs come from core-collapse SNe, then ${{dN}\over{dt}} \propto$ SFR. This is probably a fair assumption, as core-collapse SNe will preferentially occur in denser media than SNe Type Ia and will therefore be more easily observable as remnants. ${{dL_{1.4}}\over{dt}}$ is simply the time derivative of equation \ref{eqLtoT}; we then use equation \ref{eqLtoT} to write $t$ in terms of $L_{1.4}$ and find
\begin{equation} \label{eqlf}
{{dN}\over{dL_{1.4}}} \propto SFR\ E_{SN}^{1.4}\ \rho_{0}^{0.5}\ L_{1.4}^{-2.1}
\end{equation}
Therefore, this simple model predicts $\beta = -2.1$, in very good agreement with our observed $\beta = -2.07$. 

Some models for magnetic field amplification predict that the energy in the magnetic field scales as $B^{2} \propto \rho_{0}\ v_{s}^{3}$, rather than $B^{2} \propto \rho_{0}\ v_{s}^{2}$ (\citealt{Bell04}; see also \citealt{Vink08}). This would predict a steeper time dependence for luminosity:
\begin{equation}
L_{\nu} \propto E_{SN}^{1.45}\ \rho_{0}^{0.3}\ t^{-1.35}
\end{equation}
and a flatter power law index for the SNR LF:
\begin{equation}
{{dN}\over{dL_{1.4}}} \propto SFR\ E_{SN}^{1.1}\ \rho_{0}^{0.2}\ L_{1.4}^{-1.7}.
\end{equation}
A SNR LF with a power law index of $\beta$ = $-$1.7 is ruled out by the data presented here. This implies that if the rest of our assumptions are valid--- namely, that radio SNRs are in their Sedov phase and have constant fractions of their energy in CRs--- then our data is not consistent with magnetic field amplification models where $B^{2} \propto v_{s}^{3}$.
\begin{figure}
\centering
\includegraphics[width=9cm]{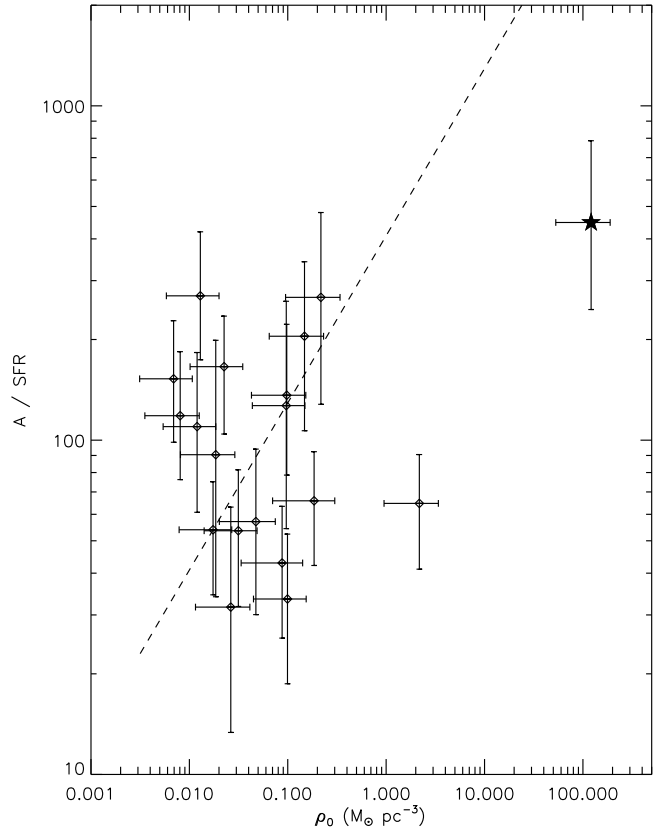}
\caption{For each galaxy, the LF scaling factor $A$ is divided by the SFR and then plotted against the ISM density. Arp 220 is marked with a star. Berezhko \& V\"{o}lk (2004) predict the dependence plotted as the dashed line: $A$/SFR $\propto \rho^{0.5}$.}
\label{aa_dens}
\end{figure}

Therefore, a magnetic field amplification model where $B^{2} \propto v_{s}^{2}$ appears to best describe our data. We have already stated that $A$ appears to be proportional to SFR; can we exclude the possibility that $A$ is actually $\propto \textrm{SFR}\  \rho_{0}^{0.5}$ as predicted in Equation \ref{eqlf}? In Figure \ref{aa_dens}, we visualize how the LF scaling factor $A$ depends on $\rho_{0}$. We divide $A$ by the SFR and plot it as a function of the density of the ISM for each galaxy; the model prediction of $A$/SFR $\propto \rho^{0.5}$ is marked with a dashed line. There does not appear to be any correlation of $A$/SFR with $\rho$ for the 18 sample galaxies. A Spearman rank correlation test gives a correlation coefficient of r$_{s}$ = $-$0.09 and a two-tailed $p$ value of 0.74, indicating no evidence of a correlation. The data are very noisy, but we note that M82 actually has a slightly lower $A$/SFR value than the dwarf irregular galaxies in our sample like IC 10 and the SMC, although the ISM density in these irregular galaxies is approximately two orders of magnitude lower than the density in M82.

``Average'' ISM density may not be a good tracer of the ISM density around SNRs, because, if all observed SNRs are from core-collapse SNe, they are exploding near their star formation sites. Therefore, the densities which are relevant for the SNR LF are those in star-forming regions, \emph{not} the global density of the galaxy. Perhaps the SNR LF is implying that the physical conditions inside star-forming regions do not vary much, even between dramatically different galaxies like the SMC and M82. This is consistent with studies of the star cluster LF \citep[e.g.,][]{Larsen02}, which imply that the masses of star clusters are relatively invariant across galaxies and unaffected by the global ISM density. 

The 18 sample galaxies form a cloud at approximately constant $A$/SFR, but Arp 220, with its extremely high ISM density, displays an unusually high value of $A$/SFR.  This is of questionable statistical significance, but may be an indication that $A$ does indeed depend on density, and that the conditions in the star-forming regions of Arp 220 are fundamentally different from those in more quiescent galaxies. More data on starburst galaxies are needed to better constrain the behavior of the SNR LF in the high $\rho_{0}$ regime, as the SNR LF may have implications for how global environment affects the physical conditions of star-forming regions. 

The lack of correlation between $A$/SFR and $\rho_{0}$ also supports our assumption that a constant fraction of SN energy goes into CR electrons. If the efficiency of cosmic ray production did vary across galaxies, the most basic expectation is that the efficiency would increase with increasing ISM density \citep{Bell78}. If we assume that the energy in cosmic rays depends on the SN explosion energy and ISM density as 
\begin{equation}
E_{CR} \propto E_{SN}\ \rho_{0}^{\gamma},
\end{equation}
where $\gamma$ is a positive scaling index, then the SNR LF of Equation \ref{eqlf} becomes modified to:
\begin{equation}
{{dN}\over{dL_{1.4}}} \propto SFR\ E_{SN}^{1.4}\ \rho_{0}^{(0.5+1.1\gamma)}\ L_{1.4}^{-2.1}.
\end{equation}
This implies that $A$ should depend even more strongly on $\rho_{0}$ than the $A \propto \rho_{0}^{0.5}$ predicted above. Of course, there is no evidence for this in the data, as we have already seen in Figure \ref{aa_dens}. The LF scaling is linearly proportional to the SFR, and there is no residual dependence on ISM density. We conclude that the density of gas surrounding SNRs does not vary much across galaxies, and therefore there is little opportunity for variable efficiency in the production of cosmic ray electrons across galaxies (regardless of the value of $\gamma$).
\section{What determines the luminosity of a galaxy's brightest SNR?}
\begin{figure}[htp]
\centering
\includegraphics[width=9cm]{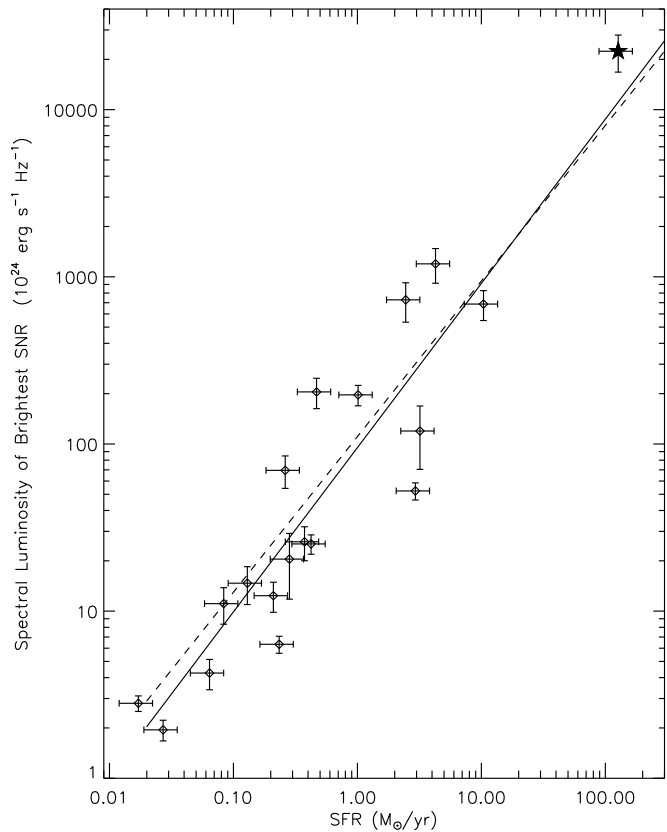}
\caption{The 1.45 GHz spectral luminosity of the brightest SNR in each galaxy plotted against SFR. The solid line represents the best linear fit in log-log space. Arp 220 is marked with a star and is excluded from the line fit. The dashed line is the mean fit to Monte Carlo simulations which represent the expected relation if only statistical sampling effects are taken into account.}
\label{sfr_maxlum}
\end{figure}
We have seen that SFR can singlehandedly account for why some galaxies have higher total 1.45 GHz spectral luminosities in their SNRs than others, but can it also explain why some galaxies host much more luminous individual remnants than others? There are two possible explanations for why the brightest remnant in NGC 253 is $\sim$3 orders of magnitude more luminous than the brightest SNR in the SMC:
\begin{enumerate}
\item The LF is actually truncated at high luminosity, and this truncation will be moved to fainter levels in galaxies with lower ISM density \citep{Hunt_Reynolds06, Thompson_etal09}. As stated above, the diameter of an SNR at the Sedov time sets the peak luminosity, and this diameter is $\propto \rho_{0}^{1/3}$. Therefore, perhaps SNRs in low-density galaxies are never as luminous as SNRs in galaxies with dense ISM. 
\item The bright end of the SNR LF is dominated by a sampling effect. In a galaxy with higher SFR, there will be a larger population of SNRs, and one is more likely to observe a SNR at its peak luminosity. In addition, there will be a larger population of extremely massive stars, and, as we expect E$_{SN}$ to correlate with the mass of the progenitor, this will increase the chance of a very luminous SNR ($L_{\nu} \propto E_{SN}^{1.3}$). Therefore, in a galaxy with a higher SFR, one has a better chance of fleshing out the high-luminosity end of the LF \citep{Lacey_Duric01}.
\end{enumerate}

The Schmidt--Kennicutt Law implies that galaxies with higher SFR will typically also have a denser ISM, so a scenario where the spectral luminosity of the brightest SNR (L$_{1.4}^{max}$) $\propto$ SFR is not in direct conflict observationally with a scenario where L$_{1.4}^{max} \propto \rho_{0}$. However, we can distinguish between these two possibilities by plotting each galaxy's L$_{1.4}^{max}$ against its SFR (Figure \ref{sfr_maxlum}). There is a solid near-linear correlation between these two quantities which is best fit with the expression:
\begin{equation}
L_{1.4}^{max} = \left(95^{+31}_{-23}\right) SFR^{0.98\pm0.12}.
\end{equation}
Again, Arp 220 is excluded from this fit.

We can test if this relation is consistent with a simple statistical sampling effect using Monte Carlo simulations. For each Monte Carlo run, we randomly sample the LF scaling $A$ from a Gaussian distribution given log $A$ = 1.966 $\pm$ 0.069 as found in Section 4.2. Next, for each sample galaxy, we calculate the number of SNRs expected given this scaling factor, an index of $\beta$ = $-$2.07, and a constant lower limit on luminosity of 0.1 $\times\ 10^{24}$ erg s$^{-1}$ Hz$^{-1}$. We randomly choose this number of SNRs from a power law probability distribution with the same lower limit on luminosity and determine the spectral luminosity of the brightest remnant (L$_{1.4}^{max,MC}$) in each galaxy. We then fit a line to log L$_{1.4}^{max,MC}$--log SFR just as was done for the real data. We perform 10$^{4}$ Monte Carlo runs in this fashion and finally calculate the mean slope and y-intercept from the 10000 individual line fits to find the relationship between L$_{max}$ and SFR which is purely due to statistical sampling:
\begin{equation}
L_{1.4}^{max,MC} = \left(111^{+46}_{-33}\right) SFR^{0.93\pm0.16}.
\end{equation}
The errors in the above equation are given by the standard deviations of the slope and y-intercept from the 10$^{4}$ line fits; they represent the scatter one might expect in the L$_{1.4}^{max}$--SFR relation due to random statistical sampling.

The observed relation is consistent with the line derived from Monte Carlo simulations to within 1$\sigma$; see also the similarity of the solid line and dashed line in Figure \ref{sfr_maxlum}. This implies that truncation at the high luminosity end is not significantly affecting the observed LF; most galaxies simply do not have enough SNe explosions to thoroughly sample the SNR LF at the luminous end (right around the Sedov time). We can not exclude the possibility that SNR LFs are truncated at high luminosities, but if they are, their cut-off luminosities are significantly higher than the brightest observed SNRs.
\section{Conclusions}
We have analyzed 20 cm SNR samples in 19 galaxies with SFRs ranging from 0.02 M$_{\sun}$ yr$^{-1}$ to 127 M$_{\sun}$ yr$^{-1}$, and we have reached the conclusion that the SNR LF is invariant across a wide range of host properties. All LFs are consistent with a power law distribution with an index $\beta \approx -$2.1. In addition, the scaling of the power law is linearly proportional to SFR, and global ISM density does not appear to affect the LF scaling. These findings are in good accordance with the model of BV04 which describes the synchrotron emission from SNRs undergoing diffusive shock acceleration and magnetic field amplification. Our data are well-matched to models of magnetic field amplification where the magnetic energy scales as $B^{2} \propto \rho_{0}\ v_{s}^{2}$, and are inconsistent with models where $B^{2} \propto \rho_{0}\ v_{s}^{3}$. In applying the models, we assume that the efficiency of CR production is constant, all SNRs are in the Sedov phase, and the densities of star-forming regions do not vary much between galaxies. These assumptions seem to describe all galaxies well, with the possible exception of the ULIRG Arp 220. Its LF scaling may be too high to be explained by SFR alone, and may imply that its star-forming regions are significantly denser than those in the other 18 sample galaxies.

In addition, we have shown that the correlation between the luminosity of a galaxy's brightest SNR and a galaxy's SFR can be completely explained by statistical sampling effects. The LF does not appear to be truncated at the high-luminosity end, and no physical justification (e.g., variations in the ISM density) is needed to explain why the brightest SNR in NGC 253 is three orders of magnitude more luminous than the brightest SNR in the SMC. Our findings support a scenario where the efficiency of CR production, the magnetic field strength in SNRs, and the density of star-forming regions are all largely independent of their host galaxies. 

\section{Acknowledgments}
 We are grateful to Ellen Zweibel, Todd Thompson, Brian Reville, John Everett and Mark Krumholz for many useful conversations. We also would like to thank Jay Strader, Tommy Nelson, and Amanda Kepley for their insights. Finally, we acknowledge the work of an anonymous referee whose comments have improved this paper. This material is based upon work supported under a National Science Foundation Graduate Research Fellowship and was also supported by NSF grant number AST-0708002.
 
This research has made use of the NASA/IPAC Extragalactic Database (NED) which is operated by the Jet Propulsion Laboratory, California Institute of Technology, under contract with the National Aeronautics and Space Administration. In addition, we acknowledge the usage of the HyperLeda database (http://leda.univ-lyon1.fr).
\bibliographystyle{apj}
\bibliography{lumfunph}
\newpage
\section*{Appendix A: Turn-Overs in Cumulative LFs at Low Luminosity}
\begin{figure*}[htp]
\centering
\includegraphics[width=12cm, angle=90]{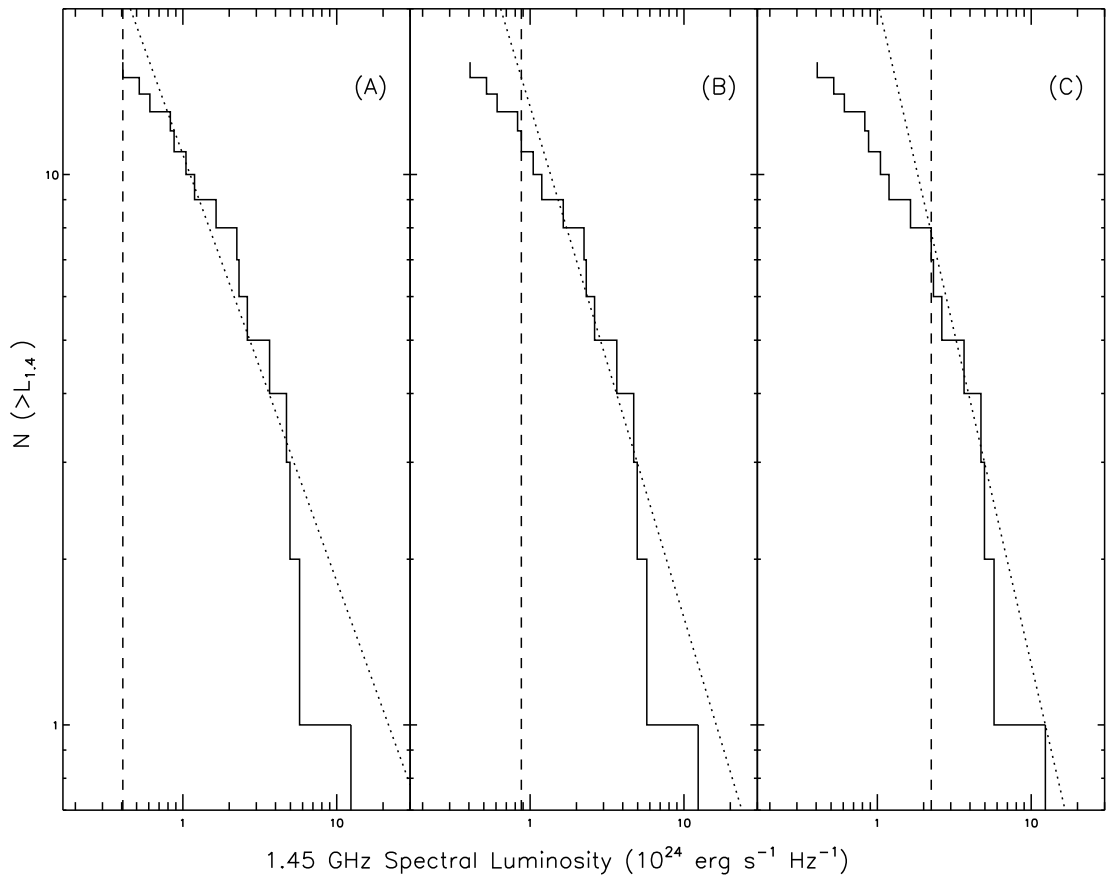}
\caption{To illustrate how the completeness limit is determined from cumulative LFs, the cumulative SNR LF for the LMC is pictured with power law fits (short-dashed lines) over three different luminosity ranges. The vertical long-dashed lines mark the trial completeness limit in each panel. In panel (A), all SNRs are included in the fit, while in panel (B), the four least-luminous SNRs (those to the left of the long-dashed line) are excluded from the fit. The fit in panel (C) excludes the eight lowest-luminosity SNRs and corresponds to the true completeness limit.}
\label{fitlmc}
\end{figure*}

As discussed in Section~\ref{completeness}, we fit a power law to the luminous end of each galaxy's cumulative LF, and the luminosity at which the cumulative LF turns over from the power law fit is considered the completeness limit. The location of this turn-over is determined iteratively through a combination of line-fitting and visual inspection. First, we attempt to fit a power law to a version of the cumulative LF which includes all SNRs in a given galaxy sample (see the left panel of Figure \ref{fitlmc}). If the power law fit approximates the cumulative LF across the entire range of luminosity, we consider the SNR sample inherently complete. However, if at the lowest luminosities the power-law fit lies systematically above the cumulative LF, this implies that the SNR sample is incomplete at these low luminosities. This occurs in the LMC, as can be seen in the left panel of Figure \ref{fitlmc}. 

Next, we try fitting a power law again, this time excluding the lowest-luminosity data point from the fit (equivalent to imposing a trial completeness limit on the data). If the cumulative LF (which now only includes data brighter than the trial completeness limit) still lies systematically below the power-law fit at low luminosities, the trial completeness limit needs to be raised to higher luminosity, and the process repeated. The central panel of Figure \ref{fitlmc} shows the fit to the LMC's cumulative LF after the four least-luminous SNRs have been excluded; we now only consider data rightward of the vertical dashed line. However, the cumulative LF directly to the right of the dashed line still lies below the power law fit, so we can see that this trial completeness limit is still too low. We continue iterating and raising the trial completeness limit until our fit to the cumulative LF looks like that in the right panel of Figure \ref{fitlmc}. The trial completeness limit has been raised to exclude the eight lowest-luminosity SNRs in the LMC, and the cumulative LF now adheres to a power law fit for luminosities brighter than the limit (rightward of the dashed line). Therefore, this is the ``true'' completeness limit for the LMC---  2.24 $\times$ 10$^{24}$ erg s$^{-1}$ Hz$^{-1}$. For all galaxies with the exception of M51, this iterative process eventually converged to reveal a similar completeness limit. In the case of M51, the most luminous remnant is so anomalously bright compared to the other SNRs that no reasonable power law fit could be achieved. We excluded the most luminous data point in M51's cumulative LF so that good power law fits were found, and then proceeded to determine the completeness limit as described above.

\LongTables
\newpage
\begin{deluxetable}{lccccccccc}
\tablecaption{ \label{tab:gals}
 Galaxy Parameters}
\tablehead{Galaxy & R.A. (2000)\tablenotemark{a}& Dec (2000)\tablenotemark{a} & Maj/Min Axis\tablenotemark{a} & i\tablenotemark{b} & Type\tablenotemark{a} & Distance\tablenotemark{c} & M$_{B}$\tablenotemark{c} & SFR\tablenotemark{d} & $\rho_{0}$\tablenotemark{e}\\
& (hr:min:sec) & ($^{\circ}$:\arcmin:\arcsec) & (arcmin) & (deg) & & (Mpc) & (mag) & (M$_{\odot}$ yr$^{-1}$) & (M$_{\odot}$ pc$^{-3}$) }
\startdata

LMC & 05:23:34.5 & $-$69:45:22 & 645 x 550 & 18 & SB(s)m & $0.05\pm0.003$ & $-$17.9 & 0.21 & 0.03 \\

SMC & 00:52:44.8 & $-$72:49:43 & 320 x 185 & 68 & SB(s)m pec & $0.06\pm0.003$ & $-$16.4 & 0.03 & 0.007\\

IC 10 & 00:20:17.3 & +59:18:14 & 6.8 x 5.9 & 31 & IBm\tablenotemark{f} & $0.66\pm0.03$ & $-$15.5 & 0.02 & 0.1 \\

M 33 & 01:33:50.9 & +30:39:36 & 70.8 x 41.7 & 55 & SA(s)cd & $0.84\pm0.04$ & $-$18.5 & 0.24 & 0.02\\

NGC 1569 & 04:30:49.0 & +64:50:53 & 3.6 x 1.8 & 65 & IBm & $1.9\pm0.4$ & $-$17.1 & 0.28 & 0.2 \\

NGC 300 & 00:54:53.5 & $-$37:41:04 & 21.9 x 15.5 & 40 & SA(s)d & $2.00\pm0.10$ & $-$17.8 & 0.10 & 0.02 \\

NGC 4214 & 12:15:39.2 & +36:19:37 & 8.5 x 6.6 & 44 & IAB(s)m & $2.92\pm0.29$ & $-$17.1 & 0.17 & 0.1 \\

NGC 2366 & 07:28:54.6 & +69:12:57 & 8.1 x 3.3 & 72 & IB(s)m & $3.19\pm0.32$ & $-$16.3 & 0.08 & 0.008 \\

M 82 & 09:55:52.7 & +69:40:46 & 11.2 x 4.3 & 79 & I0 & $3.53\pm0.35$ & $-$18.8 & 10.4 & 2 \\

M 81 & 09:55:33.2 & +69:03:55 & 26.9 x 14.1 & 59 & SA(s)ab & $3.63\pm0.18$ & $-$20.2 & 0.48 & 0.01 \\

NGC 7793 & 23:57:49.8 & $-$32:35:28 & 9.3 x 6.3 & 53 & SA(s)d & $3.91\pm0.39$ & $-$18.4 & 0.26 & 0.02 \\

NGC 253 & 00:47:33.1 & $-$25:17:18 & 27.5 x 6.8 & 78 & SAB(s)c & $3.94\pm0.39$ & $-$20.0 & 6.18 & 0.2 \\

NGC 4449 & 12:28:11.9 & +44:05:40 & 6.2 x 4.4 & 64 & IBm & $4.21\pm0.42$ & $-$18.2 & 0.47 & 0.03 \\

M 83 & 13:37:00.9 & $-$29:51:56 &12.9 x 11.5 & 46 & SAB(s)c & $4.47\pm0.22$ & $-$20.3 & 3.00 & 0.1 \\

NGC 4736 & 12:50:53.0 & +41:07:14 & 11.2 x 9.1 & 35 & (R)SA(r)ab & $4.66\pm0.47$ & $-$19.4 & 0.46 & 0.2 \\

NGC 6946 & 20:34:52.3 & +60:09:14 & 11.5 x 9.8 & 31 & SAB(rs)cd & $5.9\pm1.2$ & $-$20.8 & 3.23 & 0.09 \\

NGC 4258 & 12:18:57.5 & +47:18:14 & 18.6 x 7.2 & 72 & SAB(s)bc & $7.98\pm0.40$ & $-$20.4 & 1.28 & 0.01 \\

M 51 & 13:29:55.7 & +47:13:53 & 9.0 x 9.0 & 30 & SAbc & $8.00\pm1.04$ & $-$20.6 & 2.45 & 0.05 \\
\hline
Arp 220 & 15:34:57.1 & +23:30:11 & 1.5 x 1.2 & 57 & S? pec & 77\tablenotemark{g} & $-$20.6\tablenotemark{h} & 127\tablenotemark{i} & 100
\enddata
\tablenotetext{a}{From NED.}
\tablenotetext{b}{From HyperLeda \citep{Paturel_etal03}, except the values for the Magellanic Clouds which come from \cite{Groenewegen00} and the inclination for NGC 2366 from \cite{Hunter_etal01}.}
\tablenotetext{c}{From \cite{Kennicutt_etal08}.}
\tablenotetext{d}{SFRs are calculated using a combination of H$\alpha$ and 25 $\mu$m fluxes as calibrated by \cite{Calzetti_etal07}, and using H$\alpha$ luminosities from \cite{Kennicutt_etal08}, and 25 $\mu$m IRAS fluxes from the following references in order of preference: \cite{Sanders_etal03}, \cite{Rice_etal88}, \cite{Lisenfeld_etal07}, \cite{Moshir_etal92}.}
\tablenotetext{e}{Rough estimate of the average ISM densities in SNR survey areas, calculated as described in Section \ref{galpar}. A constant scale height is assumed for all galaxies, and therefore the uncertainty on these calculations is at least 50\%.}

\tablenotetext{f}{From \cite{deVaucouleurs_Freeman72}.}
\tablenotetext{g}{Calculated assuming H$_{0}$ = 75 km s$^{-1}$ Mpc$^{-1}$.}
\tablenotetext{h}{Calculated using the RC3 $B$ mag as provided by NED, the average of the foreground extinctions in \cite{Burstein_Heiles78} and \cite{Schlegel_etal98}, and a distance of 77 Mpc.}
\tablenotetext{i}{As determined by \cite{Anantharamaiah_etal00}, but converted to the IMF used in the \cite{Calzetti_etal07} calibration.}
\end{deluxetable} 

\newpage
\begin{deluxetable}{lccccccc}
\scriptsize
\tablecaption{ \label{tab:obs}
20 cm SNR Survey Parameters}
\tablehead{Galaxy & 20 cm Freq & Telescope/ & Synthesized Beam & Spatial Res.\tablenotemark{a} & RMS Sensitivity & F.O.V. & SFR Frac.\tablenotemark{b}\\
 & (GHz) & Config. & (arcsec$^{2}$) & (pc) & ($\mu$Jy/beam) & (arcmin) & }
\startdata
LMC\tablenotemark{d} & 1.40 & Parkes & 912 x 912 & 221 & 30000 & 600 & 1.0 \\
SMC\tablenotemark{e} & 1.42 & ATCA/375 \& Parkes & 98.0 x 98.0 & 29 & 1800 & 270 \\
IC 10\tablenotemark{f} & 1.49 & VLA/B,C & 5.0 x 5.0 & 16 & 42 & 5.5 & 0.91 \\
M33\tablenotemark{g} &1.42 & VLA/B \& WSRT & 7.0 x 7.0 & 29 & 50 & 40 & 0.98 \\
NGC 1569\tablenotemark{h} & 1.49 & VLA/A & 1.4 x 1.4 & 13 & 21 & 3\tablenotemark{c} & 1.0 \\
NGC 300\tablenotemark{i} & 1.45 & VLA/BnA & 4.7 x 3.6 & 43 & 60 & 9 & 0.65 \\ 
NGC 4214\tablenotemark{h} & 1.49 & VLA/A & 1.4 x 1.4 & 20 & 19 & 3\tablenotemark{c} & 0.77 \\
NGC 2366\tablenotemark{h} & 1.43 & VLA/A,B & 3.7 x 3.7 & 57 & 22 & 7\tablenotemark{c} & 1.0 \\
M82\tablenotemark{j} & 1.45 & VLA/A & 1.2 x 0.9 & 18 & 110 & 1 & 1.0 \\
M81\tablenotemark{k} & 1.47 & VLA/B,C & 10.0 x 10.0 & 176 & 48 & 15 & 0.88 \\
NGC 7793\tablenotemark{l} & 1.47 & VLA/BnA & 9.4 x 4.1 & 118 & 60 & 9 & 1.0 \\
NGC 253 (r$<$200 pc)\tablenotemark{m} & 1.49 & VLA/A & 1.7 x 0.9 & 24 & 60 & 1& --- \\
NGC 253 (r$>$200 pc)\tablenotemark{n} & 1.49 & VLA/A & 2.8 x 1.7 & 42 & 41 & 4\tablenotemark{c} & 0.69 \\
NGC 4449\tablenotemark{h} & 1.45 & VLA/A & 1.4 x 1.4 & 29 & 25 & 7 & 1.0 \\
M83\tablenotemark{o} & 1.45 & VLA/BnA & 3.7 x 3.7 & 79 & 74 & 9 & 0.98 \\
NGC 4736\tablenotemark{p} & 1.45 & VLA/A,B & 1.5 x 1.5 & 34 & 60 & 2 & 0.81 \\
NGC 6946\tablenotemark{q} & 1.45 & VLA/A & 1.9 x 1.6 & 50 & 16 & 9 & 0.99 \\
NGC 4258\tablenotemark{r} & 1.49 & VLA/A,B,C,D & 3.4 x 3.3 & 140 & 30 & 7\tablenotemark{c} & 0.79 \\
M51\tablenotemark{s} & 1.43 & VLA/A & 1.5 x 1.2 & 52 & 23 & 9 & 1.0 \\
\hline
Arp 220\tablenotemark{t} & 1.65 & GVLBI & 0.006 x 0.003 & 1.5 & 9 & 0.02 & 1.0
\enddata
\tablenotetext{a}{Geometric mean of the synthesized beam major and minor axes, taken at the distances listed in Table 1.}
\tablenotetext{b}{Fraction of the SFR (from Table 1) inside the SNR survey area.}
\tablenotetext{c}{Approximate: sensitivity at large radius is limited by chromatic aberration (bandwidth smearing).}
\tablenotetext{d}{\cite{Filipovic_etal98}; 
$^{e}$\cite{Filipovic_etal02} and \cite{Payne_etal04}; 
$^{f}$\cite{Yang_Skillman93}; 
$^{g}$\cite{Gordon_etal99}; 
$^{h}$\cite{Chomiuk_Wilcots09};
$^{i}$\cite{Pannuti_etal00}; 
$^{j}$\cite{Allen_Kronberg98} and \cite{McDonald_etal02}; 
$^{k}$\cite{Kaufman_etal87}; 
$^{l}$\cite{Pannuti_etal02}; 
$^{m}$\cite{Ulvestad_Antonucci97}; 
$^{n}$\cite{Ulvestad00}; 
$^{o}$\cite{Maddox_etal06}; 
$^{p}$\cite{Duric_Dittmar88}; 
$^{q}$\cite{Lacey_etal97}; 
$^{r}$\cite{Hyman_etal01}; 
$^{s}$\cite{Maddox_etal07};
$^{t}$\cite{Parra_etal07}}
\end{deluxetable} 
\\
\\
\\
\begin{deluxetable}{lcccc}
\tablecaption{ \label{tab:lmc}
  SNRs in the LMC taken from \cite{Filipovic_etal98}}
\tablehead{ID\tablenotemark{a} & R.A. (2000) & Dec. (2000) & S$_{1.4}$ & $\alpha$ \\
 & (hr:min:sec) & ($^{\circ}$:\arcmin:\arcsec) & (mJy) &    }
\startdata 
B0450-7055* & 04:49:48.85 & -70:51:08.8 & $749\pm125$ & $-0.35\pm0.16$ \\
B0453-6834 & 04:53:05.93 & -68:27:10.8 & $293\pm50$ & $-0.62\pm0.17$ \\
B0455-6843 & 04:56:24.90 & -68:45:58.8 & $277\pm47$ & $-0.57\pm0.17$ \\
B0500-7014 & 05:00:12.00 & -70:10:19.2 & $398\pm67$ & $-0.62\pm0.16$ \\
B0507-7029 & 05:07:28.15 & -70:23:53.2 & $350\pm58$ & $-0.74\pm0.17$ \\
B0509-6720 & 05:10:02.48 & -67:13:53.2 & $136\pm25$ & $-0.91\pm0.21$ \\
B0519-6941* & 05:19:05.28 & -69:36:30.8 & $1660\pm274$ & $-0.55\pm0.16$ \\
B0520-6531* & 05:21:16.91 & -65:28:38.1 & $774\pm128$ & $-0.36\pm0.16$ \\
B0521-6545 & 05:21:27.01 & -65:41:50.8 & $204\pm36$ & $-0.31\pm0.17$ \\
B0525-6941* & 05:25:12.44 & -69:38:58.2 & $4133\pm682$ & $-0.51\pm0.16$ \\
B0525-6601* & 05:25:43.01 & -65:59:01.3 & $1223\pm202$ & $-0.96\pm0.16$ \\
B0528-6551 & 05:27:40.71 & -65:49:45.8 & $174\pm31$ & $-0.40\pm0.18$ \\
B0529-6702 & 05:29:38.15 & -67:00:04.6 & $549\pm91$ & $-0.82\pm0.16$ \\
B0535-6603* & 05:35:36.11 & -66:02:31.3 & $1575\pm260$ & $-0.68\pm0.16$ \\
B0547-6942* & 05:47:00.58 & -69:39:39.1 & $1916\pm317$ & $-1.03\pm0.16$ \\
B0550-6823* & 05:50:27.71 & -68:23:00.8 & $877\pm146$ & $-0.69\pm0.16$ 
\enddata
\tablenotetext{a}{An asterisk following the SNR ID denotes that the SNR is in the complete sample for this galaxy. See Section 2.4 for more explanation.}
\end{deluxetable} 
\newpage
\begin{deluxetable}{lcccc}
\tablecaption{ \label{tab:smc}
  SNRs in the SMC taken from \cite{Payne_etal04}}
\tablehead{ID & R.A. (2000) & Dec. (2000) & S$_{1.4}$ & $\alpha$\\
 & (hr:min:sec) & ($^{\circ}$:\arcmin:\arcsec) & (mJy) & }
\startdata
J004100-733648 & 00:41:00.10 & -73:36:48.6 &  $77.6\pm7.9$ & $-0.94\pm0.14$ \\
J004637-730823 & 00:46:37.64 & -73:08:23.2 &  $67.0\pm7.2$ & $-0.61\pm0.55$ \\
J004716-730811* & 00:47:16.61 & -73:08:11.5 &  $452.2\pm45.3$ & $-0.57\pm0.11$ \\
J004728-730601 & 00:47:28.58 & -73:06:01.5 &  $137.0\pm14.6$ & $-0.36\pm0.29$ \\
J004748-731727  & 00:47:48.64 & -73:17:27.4 &  $32.5\pm3.8$ & $-0.58\pm0.04$ \\
J004806-730842 & 00:48:06.06 & -73:08:42.7 &  $82.2\pm8.5$ & $-0.22\pm0.20$ \\
J004821-731931 & 00:48:21.24 & -73:19:31.6 &  $144.6\pm14.7$ & $-1.06\pm0.11$ \\
J004907-731402* & 00:49:07.75 & -73:14:02.0 &  $277.4\pm29.2$ & $-0.62\pm1.52$ \\
J005110-732212 & 00:51:10.24 & -73:22:12.5 &  $100.3\pm11.5$ & $-0.66\pm0.04$ \\
J005240-723820* & 00:52:40.58 & -72:38:20.3 &  $211.4\pm26.9$ & $-1.03\pm0.05$ \\
J005817-721814 & 00:58:17.39 & -72:18:14.5 &  $70.2\pm10.7$ & $-0.74\pm0.10$ \\
J005927-721010* & 00:59:27.42 & -72:10:10.2 &  $363.8\pm36.9$ & $-0.81\pm0.27$ \\
J010023-713322* & 01:00:23.26 & -71:33:22.6 &  $202.6\pm20.6$ & $-0.78\pm0.06$ \\
J010313-720958 & 01:03:13.74 & -72:09:58.9 &  $94.3\pm9.6$ & $-0.47\pm0.09$ \\
J010402-720149* & 01:04:02.01 & -72:01:49.9 &  $274.2\pm27.5$ & $-0.65\pm0.02$ \\
J010505-722319 & 01:05:05.62 & -72:23:19.0 &  $92.7\pm11.6$ & $-0.68\pm0.05$ \\
J010524-720923 & 01:05:24.20 & -72:09:23.4 &  $49.1\pm6.4$ & $-0.47\pm0.02$ \\
J010539-720341 & 01:05:39.20 & -72:03:41.7 &  $48.9\pm5.2$ & $-1.89\pm0.44$ 
\enddata
\end{deluxetable} 
\begin{deluxetable}{lccccc}
\tablecaption{ \label{tab:ic10}
  SNRs in IC 10 taken from \cite{Yang_Skillman93}}
\tablehead{ID & R.A. (2000) & Dec. (2000) & S$_{1.4}$ & $\alpha$\\
 & (hr:min:sec) & ($^{\circ}$:\arcmin:\arcsec) & (mJy) & }
\startdata
HL20a* &  00:20:10.15 & +59:19:13.6 & $1.99\pm0.16$ & $-0.23\pm0.13$ \\
HL20b* & 00:20:14.99 & +59:18:53.5 & $3.40\pm0.19$ & $-0.70\pm0.12$ \\
HL50*  & 00:20:19.12 & +59:18:53.5 & $5.39\pm0.30$ & $-0.20\pm0.05$ \\
Superbubble* & 00:20:29.49 & +59:16:43.5 & $101.4\pm5.1$ & $-$0.41
\enddata
\end{deluxetable} 
\begin{deluxetable}{lccccc}
\tablecaption{ \label{tab:m33}
  SNRs in M33 taken from \cite{Gordon_etal99}}
\tablehead{ID & R.A. (2000) & Dec. (2000) & S$_{1.4}$ & $\alpha$ \\
 & (hr:min:sec) & ($^{\circ}$:\arcmin:\arcsec) & (mJy) & }
\startdata
2* &  1:32:30.49 & 30:27:42.5 & $4.9\pm0.7$ & $-0.5\pm0.1$ \\
4* &  1:32:39.83 & 30:38:17.9 & $0.7\pm0.2$ & $-0.7\pm0.3$ \\
5* &  1:32:42.30 & 30:20:54.6 & $2.3\pm0.6$ & $-0.5\pm0.3$ \\
8* &  1:32:53.29 & 30:38:10.2 & $0.9\pm0.2$ & $-1.5\pm0.8$ \\
9* &  1:32:56.49 & 30:40:39.7 & $1.0\pm0.2$ & $-0.6\pm0.2$ \\
11* & 1:32:57.01 & 30:39:25.8 & $0.7\pm0.2$ & $-0.5\pm0.3$ \\
16 & 1:33:02.59 & 30:29:38.5 & $0.4\pm0.1$ & $-1.7\pm0.8$ \\
17* & 1:33:03.52 & 30:31:20.7 & $1.1\pm0.2$ & $-0.8\pm0.3$ \\
20* & 1:33:03.90 & 30:39:54.5 & $0.6\pm0.1$ & $-0.3\pm0.3$ \\
22* & 1:33:09.27 & 30:29:51.8 & $0.8\pm0.2$ & $-0.5\pm0.2$ \\
24* & 1:33:10.04 & 30:39:33.2 & $2.3\pm0.2$ & $-0.5\pm0.1$ \\
27* & 1:33:11.00 & 30:27:42.5 & $1.0\pm0.2$ & $-0.2\pm0.2$ \\
25* & 1:33:11.03 & 30:39:44.4 & $0.8\pm0.2$ & $-0.7\pm0.3$ \\
26* & 1:33:11.17 & 30:45:16.4 & $3.1\pm0.2$ & $-0.2\pm0.1$ \\
28* & 1:33:11.23 & 30:34:22.0 & $0.7\pm0.3$ & $-0.5\pm0.5$ \\
29* & 1:33:11.96 & 30:38:47.4 & $3.5\pm0.3$ & $-0.2\pm0.1$ \\
32* & 1:33:14.60 & 30:45:14.7 & $0.9\pm0.2$ & $-0.2\pm0.2$ \\
39* & 1:33:22.60 & 30:27:04.8 & $1.1\pm0.1$ & $-0.2\pm0.1$ \\
42* & 1:33:23.80 & 30:26:12.5 & $1.4\pm0.2$ & $-0.9\pm0.2$ \\
43* & 1:33:24.76 & 30:25:32.9 & $0.6\pm0.1$ & $-0.4\pm0.3$ \\
46* & 1:33:26.80 & 30:47:43.4 & $0.8\pm0.2$ & $-1.0\pm0.5$ \\
47* & 1:33:28.06 & 30:31:34.7 & $1.5\pm0.1$ & $-0.8\pm0.2$ \\
49* & 1:33:28.89 & 30:40:25.0 & $7.0\pm0.4$ & $-0.2\pm0.0$ \\
50* & 1:33:29.03 & 30:42:17.4 & $0.8\pm0.2$ & $-0.2\pm0.2$ \\
52* & 1:33:29.47 & 30:49:10.3 & $0.5\pm0.1$ & $-0.8\pm0.5$ \\
55* & 1:33:29.99 & 30:31:47.9 & $1.8\pm0.1$ & $-0.3\pm0.1$ \\
54* & 1:33:30.12 & 30:47:35.7 & $0.5\pm0.2$ & $-0.2\pm0.4$ \\
57* & 1:33:31.22 & 30:33:33.8 & $1.8\pm0.1$ & $-0.8\pm0.1$ \\
61* & 1:33:34.88 & 30:37:06.3 & $1.4\pm0.1$ & $-0.3\pm0.1$ \\
64* & 1:33:35.90 & 30:36:28.3 & $3.6\pm0.3$ & $-0.6\pm0.1$ \\
68 & 1:33:37.65 & 30:32:00.7 & $0.4\pm0.1$ & $-0.7\pm0.3$ \\
70* & 1:33:39.05 & 30:32:37.9 & $0.8\pm0.1$ & $-0.5\pm0.2$ \\
73* & 1:33:40.49 & 30:45:57.8 & $1.0\pm0.2$ & $-0.2\pm0.1$ \\
74* & 1:33:40.73 & 30:52:13.0 & $0.5\pm0.1$ & $-1.1\pm0.7$ \\
75* & 1:33:40.74 & 30:32:22.2 & $0.5\pm0.1$ & $-0.8\pm0.3$ \\
77* & 1:33:41.67 & 30:21:03.8 & $1.3\pm0.3$ & $-0.8\pm0.3$ \\
76* & 1:33:41.74 & 30:41:50.7 & $0.9\pm0.1$ & $-0.3\pm0.2$ \\
78 & 1:33:42.37 & 30:32:58.6 & $0.3\pm0.1$ & $-0.2\pm0.4$ \\
81* & 1:33:43.44 & 30:41:02.5 & $1.8\pm0.1$ & $-0.6\pm0.1$ \\
84 & 1:33:45.00 & 30:36:00.5 & $0.3\pm0.1$ & $-1.3\pm0.8$ \\
85 & 1:33:45.42 & 30:36:26.8 & $0.4\pm0.1$ & $-0.7\pm0.3$ \\
86* & 1:33:45.46 & 30:36:49.9 & $1.1\pm0.1$ & $-0.3\pm0.1$ \\
87* & 1:33:46.95 & 30:33:36.0 & $0.5\pm0.2$ & $-0.5\pm0.4$ \\
90* & 1:33:47.88 & 30:33:05.4 & $1.0\pm0.1$ & $-0.3\pm0.2$ \\
92 & 1:33:48.33 & 30:39:35.0 & $0.3\pm0.1$ & $-0.2\pm0.2$ \\
94 & 1:33:49.59 & 30:39:54.6 & $0.4\pm0.1$ & $-0.3\pm0.2$ \\
95 & 1:33:49.93 & 30:39:42.5 & $0.3\pm0.1$ & $-0.6\pm0.3$ \\
99* & 1:33:50.22 & 30:35:28.2 & $0.6\pm0.2$ & $-0.7\pm0.5$ \\
101 & 1:33:50.67 & 30:41:19.9 & $0.4\pm0.1$ & $-0.8\pm0.3$ \\
104* & 1:33:51.74 & 30:40:55.2 & $0.5\pm0.1$ & $-0.7\pm0.3$ \\
105* & 1:33:51.75 & 30:31:01.0 & $0.7\pm0.1$ & $-0.8\pm0.3$ \\
107* & 1:33:52.74 & 30:43:51.5 & $0.6\pm0.1$ & $-0.3\pm0.2$ \\
108* & 1:33:53.23 & 30:39:05.9 & $0.5\pm0.1$ & $-0.2\pm0.2$ \\
112* & 1:33:54.86 & 30:33:10.8 & $7.5\pm0.5$ & $-0.5\pm0.1$ \\
111* & 1:33:54.97 & 30:45:21.7 & $2.1\pm0.2$ & $-0.5\pm0.1$ \\
114 & 1:33:56.32 & 30:34:53.7 & $0.4\pm0.1$ & $-0.9\pm0.7$ \\
116* & 1:33:56.96 & 30:40:48.8 & $1.1\pm0.1$ & $-1.0\pm0.2$ \\
117* & 1:33:57.12 & 30:28:50.4 & $1.1\pm0.1$ & $-0.6\pm0.2$ \\
123 & 1:33:58.46 & 30:32:19.1 & $0.3\pm0.1$ & $<-0.7$ \\
131* & 1:34:00.32 & 30:34:20.7 & $0.5\pm0.1$ & $-0.2\pm0.2$ \\
130* & 1:34:00.36 & 30:42:17.3 & $0.5\pm0.1$ & $-0.2\pm0.2$ \\
134* & 1:34:01.52 & 30:36:30.3 & $0.5\pm0.1$ & $-0.4\pm0.2$ \\
135 & 1:34:01.54 & 30:36:10.2 & $0.4\pm0.1$ & $-0.5\pm0.4$ \\
138* & 1:34:02.45 & 30:31:06.2 & $0.6\pm0.2$ & $-0.8\pm0.5$ \\
139* & 1:34:03.38 & 30:44:43.7 & $0.6\pm0.1$ & $<-0.4$ \\
140 & 1:34:04.17 & 30:32:58.4 & $0.4\pm0.1$ & $-0.5\pm0.3$ \\
143 & 1:34:06.63 & 30:48:55.3 & $0.4\pm0.1$ & $-0.6\pm0.3$ \\
147* & 1:34:09.97 & 30:31:58.0 & $1.3\pm0.1$ & $-0.3\pm0.1$ \\
148* & 1:34:10.71 & 30:42:24.2 & $1.0\pm0.1$ & $-0.4\pm0.2$ \\
154 & 1:34:15.42 & 30:33:01.2 & $0.4\pm0.1$ & $-0.4\pm0.3$ \\
157* & 1:34:16.46 & 30:51:55.3 & $6.1\pm0.4$ & $-0.2\pm0.1$ \\
158* & 1:34:16.49 & 30:52:50.1 & $1.8\pm0.2$ & $-0.4\pm0.1$ \\
159 & 1:34:16.84 & 30:39:20.3 & $0.4\pm0.1$ & $-0.5\pm0.2$ \\
160 & 1:34:17.15 & 30:41:24.4 & $0.4\pm0.1$ & $-0.5\pm0.3$ \\
168* & 1:34:29.58 & 30:41:26.4 & $0.8\pm0.1$ & $-0.4\pm0.2$ \\
170 & 1:34:30.08 & 30:35:44.6 & $0.3\pm0.1$ & $-0.9\pm0.4$ \\
180* & 1:34:39.80 & 30:41:47.8 & $1.9\pm0.2$ & $-0.2\pm0.1$ 
\enddata
\end{deluxetable} 
\newpage
\begin{deluxetable}{lcccc}
\tablecaption{ \label{tab:n1569}
  SNRs in NGC 1569 taken from \cite{Chomiuk_Wilcots09}}
\tablehead{ID & R.A. (2000) & Dec. (2000) & S$_{1.4}$ & $\alpha$ \\
 & (hr:min:sec) & ($^{\circ}$:\arcmin:\arcsec) & (mJy) & }
\startdata
N1569-04* & 4:30:44.35 &  64:51:20.1  &  $1.19\pm0.08$ & $-0.54\pm0.10$ \\
N1569-05* & 4:30:45.32 &  64:51:14.9  &  $0.71\pm0.14$ & $-0.48\pm0.26$ \\
N1569-06 & 4:30:45.60 & 64:51:15.8 & $0.09\pm0.03$ & $<-0.30$\\
N1569-07 & 4:30:45.75 &  64:51:09.6  &  $0.23\pm0.07$ & $-0.74\pm0.39$ \\
N1569-08* & 4:30:45.79 &  64:50:58.2  &  $0.76\pm0.09$ & $-0.68\pm0.12$ \\
N1569-09 & 4:30:46.03 &  64:51:08.0  &  $0.30\pm0.08$ & $-0.60\pm0.11$ \\
N1569-11* & 4:30:46.54 &  64:50:53.6  &  $1.03\pm0.09$ & $ -0.55\pm0.13$ \\ 
N1569-12 & 4:30:46.83 &  64:50:37.8  &  $0.28\pm0.08$ & $<-0.34$ \\
N1569-14* & 4:30:46.97 &  64:51:07.2  &  $1.33\pm0.11$ & $-0.76\pm0.14$ \\
N1569-16* & 4:30:47.48 &  64:50:55.7  &  $0.50\pm0.11$ & $-0.27\pm0.19$ \\
N1569-17* & 4:30:47.73 &  64:51:09.7  &  $0.89\pm0.17$ & $-0.29\pm0.29$ \\
N1569-18 & 4:30:47.91 &  64:50:50.3  &  $0.37\pm0.12$ & $-0.20\pm0.25$ \\
N1569-20 & 4:30:48.20 &  64:50:54.7  &  $0.27\pm0.08$ & $-0.31\pm0.12$ \\
N1569-21* & 4:30:48.42 &  64:50:53.6  &  $0.83\pm0.17$ & $-0.26\pm0.27$ \\
N1569-23* & 4:30:48.48 &  64:51:08.5  &  $0.47\pm0.12$ & $-0.21\pm0.21$ \\
N1569-27* & 4:30:49.50 &  64:50:59.4  &  $1.47\pm0.24$ & $-0.28\pm0.39$ \\
N1569-28* & 4:30:51.55 &  64:50:51.0  &  $0.57\pm0.08$ & $-0.67\pm0.12$ \\
N1569-30 & 4:30:52.04 &  64:50:44.4  &  $0.40\pm0.10$ & $-0.23\pm0.15$ \\
N1569-32* & 4:30:52.19 & 64:50:54.8 & $0.57\pm0.17$ & $<-0.41$\\
N1569-33* & 4:30:52.42 &  64:50:43.2  &  $0.50\pm0.10$ & $-0.59\pm0.31$ \\
N1569-34 & 4:30:52.46 &  64:50:51.6  &  $0.29\pm0.08$ & $<-0.44$ \\
N1569-35 & 4:30:52.96 &  64:50:48.8  &  $0.25\pm0.08$ & $-0.25\pm0.12$ \\
N1569-36 & 4:30:53.36 &  64:50:44.5  &  $0.22\pm0.04$ & $-0.29\pm0.20$ \\
N1569-37 & 4:30:53.53 &  64:50:47.7  &  $0.32\pm0.10$ & $-0.46\pm0.15$ \\
N1569-38* & 4:30:54.08 &  64:50:43.5  &  $4.74\pm0.25$ & $-0.58\pm0.15$ 
\enddata
\end{deluxetable} 
\begin{deluxetable}{lccccc}
\tablecaption{ \label{tab:n300}
  SNRs in NGC 300 taken from \cite{Pannuti_etal00}}
\tablehead{ID & R.A. (2000) & Dec. (2000) & S$_{1.4}$ & $\alpha$ \\
 & (hr:min:sec) & ($^{\circ}$:\arcmin:\arcsec) & (mJy) & }
\startdata
R1* & 00:54:38.2 & -37:41:47 & $0.36\pm0.07$ & $<-1.1$ \\
R2* & 00:54:38.4 & -37:42:42 & $0.60\pm0.14$ & $<-1.5$ \\
R3* & 00:54:43.4 & -37:43:11 & $0.57\pm0.10$ & $-0.6\pm0.2$ \\
R4* & 00:54:44.9 & -37:41:10 & $0.24\pm0.07$ & $<-0.7$ \\
R5* & 00:54:45.1 & -37:41:49 & $0.23\pm0.07$ & $<-0.7$ \\
R6* & 00:54:50.3 & -37:40:31 & $0.30\pm0.10$ & $-0.6\pm0.4$ \\
R7* & 00:54:51.1 & -37:40:59 & $0.22\pm0.07$ & $<-0.7$ \\
R8* & 00:54:51.1 & -37:41:45 & $0.30\pm0.10$ & $<-0.9$ \\
R9 & 00:54:51.3 & -37:46:22 & $0.24\pm0.07$ & $<-0.7$ \\
R10* & 00:54:51.8 & -37:39:39 & $0.48\pm0.09$ & $-0.4\pm0.3$ \\
R11* & 00:55:03.6 & -37:42:49 & $0.35\pm0.11$ & $-0.2\pm0.3$ \\
R12* & 00:55:03.7 & -37:43:21 & $0.42\pm0.10$ & $-1.0\pm0.6$ \\
R13* & 00:55:12.6 & -37:41:38 & $0.50\pm0.14$ & $-0.4\pm0.3$ \\
R14* & 00:55:30.1 & -37:39:20 & $0.28\pm0.07$ & $<-0.9$ \\
N300-S10* & 00:54:40.6 & -37:40:54 & $0.29\pm0.07$ & $-0.6\pm0.3$ \\
N300-S11* & 00:54:43.4 & -37:43:10 & $0.89\pm0.16$ & $-0.7\pm0.2$ \\
N300-S26* & 00:55:15.5 & -37:44:41 & $0.22\pm0.07$ & $<-0.7$ 
\enddata
\end{deluxetable} 
\begin{deluxetable}{lccccc}
\tablecaption{ \label{tab:n4214}
  SNRs in NGC 4214 taken from \cite{Chomiuk_Wilcots09}}
\tablehead{ID & R.A. (2000) & Dec. (2000) & S$_{1.4}$ & $\alpha$ \\
 & (hr:min:sec) & ($^{\circ}$:\arcmin:\arcsec) & (mJy) & }
\startdata
N4214-02*  & 12:15:34.74 & 36:20:17.1 & $0.48\pm0.10$ & $-0.48\pm0.14$ \\
N4214-03  & 12:15:38.18 & 36:19:44.9 & $0.18\pm0.05$ & $<-0.20$\\
N4214-04  & 12:15:38.98 & 36:18:59.1 & $0.28\pm0.06$ & $-0.48\pm0.31$ \\
N4214-08  & 12:15:39.78 & 36:19:34.3 & $0.16\pm0.04$ & $<-0.28$\\
N4214-09* & 12:15:39.99 & 36:18:41.1 & $0.47\pm0.05$ & $-0.62\pm0.07$ \\
N4214-10* & 12:15:40.02 & 36:19:35.5 & $1.44\pm0.23$ & $<-0.37$ \\
N4214-11* & 12:15:40.12 & 36:19:30.7 & $0.67\pm0.11$ & $-0.53\pm0.17$ \\
N4214-12* & 12:15:40.55 & 36:19:31.5 & $1.08\pm0.17$ & $-0.56\pm0.26$ \\
N4214-18  & 12:15:41.64 & 36:19:09.7 & $0.15\pm0.05$ & $<-0.20$\\
N4214-19 & 12:15:41.87 & 36:19:15.3 & $0.30\pm0.04$ & $<-0.43$ 
\enddata
\end{deluxetable} 
\begin{deluxetable}{lccccc}
\tablecaption{ \label{tab:n2366}
  SNRs in NGC 2366 taken from \cite{Chomiuk_Wilcots09}}
\tablehead{ID & R.A. (2000) & Dec. (2000) & S$_{1.4}$ & $\alpha$ \\
 & (hr:min:sec) & ($^{\circ}$:\arcmin:\arcsec) & (mJy) & }
\startdata
N2366-07* & 7:28:30.41 & 69:11:33.8 & $0.20\pm0.03$ & $-0.53\pm0.04$ \\
N2366-12* & 7:28:45.26 & 69:12:19.8 & $0.91\pm0.13$ & $-0.91\pm0.21$ \\
N2366-15* & 7:28:52.10 & 69:12:54.4 & $0.19\pm0.05$ & $-0.37\pm0.07$ \\
N2366-16* & 7:28:54.57 & 69:11:12.7 & $0.15\pm0.03$ & $-0.29\pm0.04$ \\
N2366-18* & 7:28:57.67 & 69:13:41.0 & $0.19\pm0.04$ & $-0.42\pm0.21$ 
\enddata
\end{deluxetable} 
\begin{deluxetable}{lccccc}
\tablecaption{ \label{tab:m82}
  SNRs in M82 taken from \cite{Allen_Kronberg98} and \cite{McDonald_etal02}}
\tablehead{ID & R.A. (2000) & Dec. (2000) & S$_{1.4}$ & $\alpha$ \\
 & (hr:min:sec) & ($^{\circ}$:\arcmin:\arcsec) & (mJy) & }
\startdata 
39.10+57.4* & 09:55:47.88 & +69:40:43.6 & $7.49\pm0.37$ & $-0.39\pm0.03$ \\
39.40+56.1* & 09:55:48.16 & +69:40:42.5 & $2.83\pm0.57$ & $-0.21\pm0.10$ \\
39.64+53.4* & 09:55:48.40 & +69:40:39.7 & $3.32\pm1.04$ & $-0.68\pm0.16$ \\
39.77+56.9 & 09:55:48.54 & +69:40:43.2 & $1.77\pm0.31$ & $-0.48\pm0.10$ \\
40.32+55.1 & 09:55:49.08 & +69:40:41.5 & $1.86\pm0.62$ & $-0.54\pm0.17$ \\
40.62+56.0* & 09:55:49.35 & +69:40:42.4 & $3.08\pm0.81$ & $-0.76\pm0.27$ \\
40.66+55.2* & 09:55:49.43 & +69:40:41.4 & $14.23\pm0.94$ & $-0.50\pm0.03$ \\
41.29+59.7* & 09:55:50.06 & +69:40:45.9 & $6.29\pm1.04$ & $-0.48\pm0.08$ \\
42.53+61.9* & 09:55:51.27 & +69:40:48.1 & $12.89\pm1.91$ & $-1.64\pm0.09$ \\
42.67+55.6* & 09:55:51.40 & +69:40:41.7 & $3.49\pm1.28$ & $-0.60\pm0.19$ \\
42.66+56.4 & 09:55:51.41 & +69:40:42.6 & $1.22\pm0.55$ & $-0.55\pm0.41$ \\
42.82+61.3* & 09:55:51.55 & +69:40:47.5 & $8.86\pm1.30$ & $-1.44\pm0.37$ \\
43.18+58.3* & 09:55:51.92 & +69:40:44.5 & $11.26\pm0.61$ & $-0.63\pm0.03$ \\
43.31+59.2* & 09:55:52.04 & +69:40:45.4 & $23.62\pm0.61$ & $-0.63\pm0.01$ \\
44.01+59.6* & 09:55:52.73 & +69:40:45.7 & $46.08\pm2.00$ & $-0.47\pm0.02$ \\
44.29+59.3* & 09:55:53.00 & +69:40:45.4 & $5.41\pm0.74$ & $-0.56\pm0.07$ \\
44.52+58.1* & 09:55:53.23 & +69:40:44.3 & $5.07\pm0.35$ & $-0.53\pm0.04$ \\
44.91+61.1* & 09:55:53.62 & +69:40:47.3 & $4.26\pm0.76$ & $-1.02\pm0.30$ \\
45.17+61.2* & 09:55:53.89 & +69:40:47.3 & $20.59\pm0.40$ & $-0.73\pm0.01$ \\
45.26+65.3* & 09:55:53.97 & +69:40:51.3 & $4.32\pm1.00$ & $-0.62\pm0.12$ \\
45.44+67.3* & 09:55:54.14 & +69:40:53.5 & $3.80\pm0.28$ & $-0.57\pm0.04$ \\
45.79+65.2* & 09:55:54.47 & +69:40:51.4 & $4.79\pm0.66$ & $-0.24\pm0.07$ \\
45.91+63.8* & 09:55:54.61 & +69:40:49.9 & $3.19\pm0.41$ & $-0.49\pm0.07$ \\
46.52+63.8* & 09:55:55.22 & +69:40:50.0 & $6.91\pm0.36$ & $-0.67\pm0.03$ \\
46.56+73.8 & 09:55:55.28 & +69:40:59.8 & $2.60\pm0.41$ & $-0.74\pm0.10$ \\
46.75+67.0* & 09:55:55.45 & +69:40:53.0 & $3.94\pm0.46$ & $-0.53\pm0.06$ \\
47.37+68.0* & 09:55:56.07 & +69:40:54.0 & $2.88\pm0.43$ & $-0.70\pm0.11$ 
\enddata
\end{deluxetable} 
\begin{deluxetable}{lccccc}
\tablecaption{ \label{tab:m81}
  SNRs in M81 taken from \cite{Kaufman_etal87}}
\tablehead{ID & R.A. (2000) & Dec. (2000) & S$_{1.4}$ & $\alpha$ \\
 & (hr:min:sec) & ($^{\circ}$:\arcmin:\arcsec) & (mJy) & }
\startdata 
178* & 09:55:47.74 & +68:59:18.3 & $1.46\pm0.13$ & $-0.30\pm0.10$ \\
181* & 09:55:52.57 & +68:59:09.1 & $1.60\pm0.14$ & $-0.31\pm0.10$ \\
221 & 09:55:52.59 & +69:05:35.1 & $0.24\pm0.08$ & $<-0.2$ \\
187* & 09:56:01.97 & +68:59:06.7 & $0.63\pm0.11$ & $-0.28\pm0.22$ \\
198* & 09:56:17.06 & +69:06:02.1 & $0.53\pm0.10$ & $<-0.4$ \\
101* & 09:56:18.62 & +69:04:29.1 & $0.62\pm0.11$ & $<-0.7$ \\
104 & 09:56:20.24 & +69:01:12.0 & $0.31\pm0.08$ & $-0.55\pm0.34$ 
\enddata
\end{deluxetable} 
\newpage
\begin{deluxetable}{lccccc}
\tablecaption{ \label{tab:n7793}
  SNRs in NGC 7793 taken from \cite{Pannuti_etal02}}
\tablehead{ID & R.A. (2000) & Dec. (2000) & S$_{1.4}$ & $\alpha$ \\
 & (hr:min:sec) & ($^{\circ}$:\arcmin:\arcsec) & (mJy) & }
\startdata
N7793-R1* & 23:57:40.2 & -32:37:38 & $0.22\pm0.07$ & $<-0.7$ \\
N7793-R2* & 23:57:41.2 & -32:34:50 & $0.39\pm0.13$ & $-0.41\pm0.40$ \\
N7793-S11* & 23:57:47.3 & -32:35:23 & $0.45\pm0.15$ & $<-0.3$ \\
N7793-R4* & 23:57:48.4 & -32:36:15 & $0.22\pm0.07$ & $-0.27\pm0.34$ \\
N7793-S26* & 23:58:00.0 & -32:33:19 & $3.80\pm0.35$ & $-0.93\pm0.16$ \\
N7793-R5* & 23:58:00.7 & -32:35:06 & $0.25\pm0.07$ & $-0.61\pm0.37$ 
\enddata
\end{deluxetable} 
\begin{deluxetable}{lccccc}
\tablecaption{ \label{tab:n253}
  SNRs in NGC 253 taken from \cite{Ulvestad_Antonucci97} and \cite{Ulvestad00}}
\tablehead{ID & R.A. (2000) & Dec. (2000) & S$_{1.4}$ & $\alpha$ \\
 & (hr:min:sec) & ($^{\circ}$:\arcmin:\arcsec) & (mJy) & }
\startdata 
1 &  00:47:26.81 & -25:17:37.2 & $0.6\pm0.1$ & $-0.34\pm0.25$ \\
2 &  00:47:26.91 & -25:17:33.0 & $0.7\pm0.1$ & $-0.72\pm0.31$ \\
3 & 00:47:27.53 & -25:17:59.0 & $0.6\pm0.1$ & $<-0.5$ \\
4 & 00:47:27.99 & -25:17:15.4 & $2.4\pm0.2$ & $-1.01\pm0.25$ \\
7 & 00:47:28.42 & -25:17:09.4 & $1.8\pm0.2$ & $-0.34\pm0.12$ \\
9 & 00:47:29.90 & -25:17:38.6 & $3.6\pm0.2$ & $-0.91\pm0.09$ \\
11 & 00:47:30.75 & -25:16:38.2 & $0.9\pm0.1$ & $-0.69\pm0.23$ \\
13 & 00:47:31.64 & -25:16:25.2  & $0.5\pm0.1$ & $-0.43\pm0.33$ \\
4.81-43.6* & 00:47:32.21 & -25:17:21.8 & $5.78\pm3.72$ & $-1.18\pm0.52$ \\
5.48-43.3* & 00:47:32.88 & -25:17:21.5 & $64.35\pm8.08$ & $-0.69\pm0.05$ \\
5.62-41.3* & 00:47:33.02 & -25:17:19.5 & $12.53\pm2.61$ & $-0.26\pm0.08$ \\
5.75-41.8* & 00:47:33.15 & -25:17:20.0 & $15.19\pm1.64$ & $-0.66\pm0.04$ \\
5.79-41.1* & 00:47:33.19 & -25:17:19.3 & $16.72\pm4.57$ & $-0.97\pm0.11$ \\
5.87-40.1* & 00:47:33.27 & -25:17:18.3 & $12.78\pm2.49$ & $-1.0\pm0.08$ \\
5.94-40.1 & 00:47:33.34 & -25:17:18.3 & $1.25\pm0.85$ & $-0.47\pm0.51$ \\
5.95-37.7 & 00:47:33.35 & -25:17:15.9 & $2.68\pm1.26$ & $-0.66\pm0.36$ \\
6.00-37.0* & 00:47:33.40 & -25:17:15.2 & $13.99\pm1.50$ & $-0.76\pm0.04$ \\
14 & 00:47:33.40 & -25:16:07.4  & $0.4\pm0.1$ & $<-0.2$ \\
6.40-37.1* & 00:47:33.79 & -25:17:15.3 & $7.03\pm1.58$ & $-0.74\pm0.17$ \\
7.60-27.7 & 00:47:34.99 & -25:17:05.9 & $0.97\pm0.66$ & $-0.33\pm0.50$ \\
18 & 00:47:37.19 & -25:16:45.0 & $4.0\pm0.2$ & $-0.81\pm0.07$ \\
19 & 00:47:39.46 & -25:16:34.3 & $2.5\pm0.3$ & $<-1.1$ \\
20* & 00:47:40.52 & -25:16:39.3 & $6.2\pm0.3$ & $-0.69\pm0.08$ \\
22 & 00:47:43.23 & -25:15:39.9 & $1.2\pm0.2$ & $-0.24\pm0.23$
\enddata
\end{deluxetable} 
\begin{deluxetable}{lccccc}
\tablecaption{ \label{tab:n4449}
  SNRs in NGC 4449 taken from \cite{Chomiuk_Wilcots09}}
\tablehead{ID & R.A. (2000) & Dec. (2000) & S$_{1.4}$ & $\alpha$ \\
 & (hr:min:sec) & ($^{\circ}$:\arcmin:\arcsec) & (mJy) & }
\startdata      
N4449-07* & 12:28:09.67 & 44:05:19.8 & $0.24\pm0.03$ & $-0.51\pm0.04$ \\
N4449-11* & 12:28:47.49 & 44:00:11.3 & $0.13\pm0.04$ & $<-0.35$ \\
N4449-12* & 12:28:10.93 & 44:06:48.8 & $9.66\pm0.49$ & $-0.90\pm0.07$ \\
N4449-14* & 12:28:11.47 & 44:05:38.6 & $1.46\pm0.17$ & $-0.54\pm0.19$ \\
N4449-17* & 12:28:12.77 & 44:06:12.2 & $0.25\pm0.06$ & $-0.78\pm0.08$ \\
N4449-19* & 12:28:13.07 & 44:05:37.8 & $0.23\pm0.06$ & $<-0.55$ \\
N4440-24* & 12:28:16.13 & 44:06:43.3 & $0.37\pm0.10$ & $-0.53\pm0.13$ \\
N4449-26* & 12:28:19.23 & 44:06:55.9 & $0.16\pm0.02$ & $-0.26\pm0.12$ 
\enddata
\end{deluxetable} 
\newpage
\begin{deluxetable}{lcccc}
\tablecaption{ \label{tab:m83}
  SNRs in M83 taken from \cite{Maddox_etal06}}
\tablehead{ID & R.A. (2000) & Dec. (2000) & S$_{1.4}$ & $\alpha$ \\
 & (hr:min:sec) & ($^{\circ}$:\arcmin:\arcsec) & (mJy) &	}
\startdata
2* &  13:36:50.83 &  -29:51:59.6 & $0.38\pm0.08$ & $<-0.7$ \\
3* &  13:36:50.86 &  -29:52:38.5 & $0.38\pm0.08$ & $-0.35\pm0.21$ \\
4* &  13:36:51.11 &  -29:50:42.0 & $0.55\pm0.08$ & $-0.72\pm0.20$ \\
7* &  13:36:52.78 &  -29:52:31.6 & $0.52\pm0.08$ & $-0.60\pm0.19$ \\
9* &  13:36:52.83 &  -29:51:38.0 & $0.62\pm0.09$ & $<-1.1$ \\
10* &  13:36:52.91 &  -29:52:49.1 & $0.60\pm0.09$ & $-0.63\pm0.19$ \\
18* &  13:36:54.74 &  -29:52:56.8 & $0.60\pm0.09$ & $-0.49\pm0.18$ \\
21 &  13:36:55.41 &  -29:52:56.1 & $0.26\pm0.08$ & $<-0.4$ \\
23* &  13:36:55.72 &  -29:49:52.1 & $0.70\pm0.09$ & $-0.40\pm0.13$ \\
24* &  13:36:56.13 &  -29:52:55.0 & $0.45\pm0.08$ & $-0.31\pm0.19$ \\
31* &  13:36:59.98 &  -29:52:16.7 & $2.19\pm0.14$ & $-0.34\pm0.07$ \\
32* &  13:37:00.17 &  -29:51:40.0 & $2.16\pm0.13$ & $-1.10\pm0.08$ \\
35* &  13:37:02.36 &  -29:51:25.9 & $0.77\pm0.09$ & $-0.39\pm0.12$ \\
41* &  13:37:06.61 &  -29:53:32.3 & $0.59\pm0.09$ & $-0.28\pm0.16$ \\
48* &  13:37:07.89 &  -29:51:17.8 & $0.98\pm0.09$ & $-0.51\pm0.10$ \\
49 &  13:37:08.09 &  -29:52:55.8 & $0.34\pm0.09$ & $-0.32\pm0.29$ \\
52* &  13:37:09.19 &  -29:51:33.3 & $0.48\pm0.07$ & $-0.31\pm 0.17$ 
\enddata
\end{deluxetable} 
\begin{deluxetable}{lccccc}
\tablecaption{ \label{tab:n4736}
  SNRs in NGC 4736 taken from \cite{Duric_Dittmar88}}
\tablehead{ID & R.A. (2000) & Dec. (2000) & S$_{1.4}$ & $\alpha$ \\
 & (hr:min:sec) & ($^{\circ}$:\arcmin:\arcsec) & (mJy) &  }
\startdata 
6 & 12:50:49.2 & +41:07:15 & $0.28\pm0.07$ & $-0.20\pm0.31$ \\
5* & 12:50:49.4 & +41:07:07 & $0.60\pm0.10$ & $-0.79\pm0.26$ \\
9 & 12:50:49.5 & +41:07:47 & $0.32\pm0.08$ & $-0.31\pm0.31$ \\
4* & 12:50:49.8 & +41:06:55 & $0.49\pm0.07$ & $-0.56\pm0.21$ \\
16  & 12:50:50.5 & +41:07:38 & $0.35\pm0.06$ & $-0.55\pm0.32$ \\
3* & 12:50:50.9 & +41:06:50 & $0.48\pm0.07$ & $-0.26\pm0.21$ \\
15* & 12:50:51.9 & +41:07:16 & $0.44\pm0.06$ & $-0.44\pm0.26$ \\
14* & 12:50:52.4 & +41:07:01 & $0.45\pm0.06$ & $-0.42\pm0.25$ \\
13* & 12:50:52.6 & +41:07:09 & $0.61\pm0.08$ & $-0.67\pm0.22$ \\
2* & 12:50:53.8 & +41:06:30 & $0.61\pm0.09$ & $-0.62\pm0.21$ \\
1* & 12:50:54.4 & +41:06:34 & $1.00\pm0.11$ & $-0.49\pm0.12$ 
\enddata
\end{deluxetable} 
\begin{deluxetable}{lcccc}
\tablecaption{ \label{tab:n6946}
  SNRs in NGC 6946 taken from \cite{Lacey_etal97}}
\tablehead{ID & R.A. (2000) & Dec. (2000) & S$_{1.4}$ & $\alpha$ \\
 & (hr:min:sec) & ($^{\circ}$:\arcmin:\arcsec) & (mJy) &	}
\startdata
8* & 20:34:31.05 & +60:08:27.5 & $0.20\pm0.03$ & $-0.58\pm0.29$  \\
12 & 20:34:34.78 & +60:11:33.6 & $0.18\pm0.04$ & $-0.58\pm0.34$  \\
13* & 20:34:34.88 & +60:11:38.5 & $0.63\pm0.08$ & $-0.38\pm0.17$ \\
19* & 20:34:36.08 & +60:11:40.4 & $0.21\pm0.04$ & $<-0.2$ \\
16* & 20:34:36.09 & +60:08:16.8 & $0.32\pm0.04$ & $-0.31\pm0.19$  \\
22* & 20:34:37.52 & +60:09:36.6 & $0.41\pm0.05$ & $-0.96\pm0.23$  \\
23* & 20:34:39.41 & +60:04:52.8 & $0.35\pm0.04$ & $-0.55\pm0.30$  \\
25 & 20:34:40.18 & +60:08:50.4 & $0.16\pm0.04$ & $<-0.4$ \\
26* & 20:34:41.39 & +60:08:46.3 & $0.87\pm0.07$ & $-0.45\pm0.11$  \\
38 & 20:34:48.40 &  +60:10:54.0 & $0.15\pm0.03$ & $<-0.4$  \\
44 & 20:34:49.69 & +60:12:40.4 & $0.17\pm0.03$ & $<-0.2$  \\
47 & 20:34:50.80 & +60:07:47.9 & $0.09\pm0.02$ & $<-0.3$  \\
48* & 20:34:50.94 & +60:10:20.6 & $0.43\pm0.06$ & $-0.36\pm0.16$  \\
50* & 20:34:51.27 & +60:09:53.7 & $0.23\pm0.07$ & $<-0.7$  \\
51* & 20:34:51.41 & +60:07:39.2 & $0.41\pm0.08$ & $-0.84\pm0.24$  \\
59* & 20:34:52.73 & +60:08:59.4 & $0.22\pm0.04$ & $-0.38\pm0.29$  \\
60 & 20:34:52.80 & +60:07:54.2 & $0.15\pm0.03$ & $-0.52\pm0.28$  \\
62* & 20:34:52.84 & +60:08:51.7 & $0.51\pm0.06$ & $-0.25\pm0.15$  \\
63* & 20:34:53.15 & +60:08:47.8 & $0.32\pm0.05$ & $<-1.0$  \\
69* & 20:34:53.72 & +60:09:18.8 & $0.64\pm0.17$ & $-0.48\pm0.31$  \\
70* & 20:34:53.85 & +60:10:29.7 & $0.21\pm0.04$ & $-0.34\pm0.25$  \\
80 & 20:34:56.69 & +60:08:26.3 & $0.14\pm0.03$ & $-0.28\pm0.25$  \\
83* & 20:34:58.65 & +60:10:51.9 & $0.51\pm0.07$ &  $-0.53\pm0.22$  \\
85* & 20:35:00.72 & +60:11:30.6 & $1.59\pm0.09$ & $-0.55\pm0.08$  \\
86 & 20:35:03.17 & +60:10:55.9 & $0.13\pm0.03$ & $<-0.6$  \\
88 & 20:35:04.04 & +60:09:54.3 & $0.11\pm0.03$ & $<-0.5$  \\
89* & 20:35:04.18 & +60:10:54.7 & $0.36\pm0.06$ & $-0.85\pm0.24$  \\
97* & 20:35:06.09 & +60:10:56.3 & $0.45\pm0.07$ & $-0.34\pm0.22$  \\
99* & 20:35:06.67 & +60:11:11.0 & $0.27\pm0.06$ & $-0.49\pm0.30$ \\
101* & 20:35:08.08 & +60:11:13.1 & $0.70\pm0.07$ & $-0.65\pm0.12$  \\
102 & 20:35:08.26 & +60:09:53.6 & $0.13\pm0.04$ & $<-0.3$  \\
103 & 20:35:08.89 & +60:09:23.7 & $0.17\pm0.05$ & $<-0.5$  \\
105* & 20:35:11.24 & +60:10:34.5 & $0.24\pm0.03$ & $-0.24\pm0.22$  \\
106* & 20:35:11.44 & +60:10:31.6 & $0.24\pm0.04$ & $-0.45\pm0.28$  \\
107* & 20:35:11.52 & +60:09:12.0 & $0.20\pm0.03$ & $<-0.4$ \\
115* & 20:35:23.65 & +60:09:50.1 & $0.34\pm0.05$ & $-0.33\pm0.26$ \\
118* & 20:35:25.25 & +60:09:57.9 & $2.87\pm0.16$ & $-0.37\pm0.08$  
\enddata
\end{deluxetable}

\begin{deluxetable}{lcccc}
\tablecaption{ \label{tab:n4258}
  SNRs in NGC 4258 taken from \cite{Hyman_etal01}}
\tablehead{ID & R.A. (2000) & Dec. (2000) & S$_{1.4}$ & $\alpha$ \\
 & (hr:min:sec) & ($^{\circ}$:\arcmin:\arcsec) & (mJy) &	}
\startdata
12* &  12:18:56.3 &  47:16:50 & $0.65\pm0.12$ & $-0.61\pm0.30$ \\
3*  & 12:18:56.5 &  47:20:14 & $0.81\pm0.14$ & $-0.20\pm0.21$ \\
4*  & 12:18:57.4  & 47:20:05  & $0.66\pm0.11$ & $-0.23\pm0.20$ \\
11* &  12:18:57.6 &  47:16:07 & $0.54\pm0.08$ & $-0.56\pm0.31$ \\
9*  & 12:19:01.3 &  47:15:25  & $2.58\pm0.25$ & $-0.25\pm0.11$ 
\enddata
\end{deluxetable}

\begin{deluxetable}{lcccc}
\tablecaption{ \label{tab:m51}
  SNRs in M51 taken from \cite{Maddox_etal07}}
\tablehead{ID & R.A. (2000) & Dec. (2000) & S$_{1.4}$ & $\alpha$ \\
 & (hr:min:sec) & ($^{\circ}$:\arcmin:\arcsec) & (mJy) &	}
\startdata
2    & 13:29:36.56  & 47:11:05.5  & $0.07\pm0.02$ & $-0.38\pm0.37$ \\
15*  & 13:29:49.60  & 47:13:27.5  & $0.32\pm0.03$ & $-1.24\pm0.16$ \\
18*  & 13:29:49.93  & 47:11:31.1  & $0.46\pm0.03$ & $-0.57\pm0.08$ \\
19*  & 13:29:49.95  & 47:11:26.7  & $0.23\pm0.02$ & $-0.59\pm0.12$ \\
20*  & 13:29:50.04  & 47:11:24.9  & $0.21\pm0.02$ & $<-1.1$ \\
22*  & 13:29:50.13  & 47:11:36.9  & $0.11\pm0.02$ & $-0.49\pm0.21$ \\
23*  & 13:29:50.20  & 47:11:51.4  & $0.15\pm0.02$ & $-0.28\pm0.15$ \\
31*  & 13:29:50.45  & 47:11:27.0  & $0.12\pm0.02$ & $-0.37\pm0.19$ \\
36*  & 13:29:51.50  & 47:12:00.5  & $0.21\pm0.02$ & $-0.21\pm0.12$ \\
40*  & 13:29:51.80  & 47:11:40.4  & $0.38\pm0.03$ & $<-2.0$ \\
41*  & 13:29:51.86  & 47:11:37.0  & $0.21\pm0.02$ & $-0.49\pm0.12$ \\
42*  & 13:29:51.99  & 47:10:54.0  & $0.12\pm0.02$ & $-0.65\pm0.23$ \\
47*  & 13:29:52.08  & 47:11:26.8  & $0.08\pm0.02$ & $-0.60\pm0.34$ \\
48*  & 13:29:52.17  & 47:11:36.6  & $0.09\pm0.02$ & $-0.21\pm0.21$ \\
49*  & 13:29:52.22  & 47:11:29.5  & $0.09\pm0.03$ & $-0.45\pm0.35$ \\
50*  & 13:29:52.35  & 47:11:36.1  & $0.09\pm0.02$ & $-0.38\pm0.23$ \\
52*  & 13:29:52.73  & 47:11:21.2  & $0.16\pm0.03$ & $-0.82\pm0.23$ \\
55*  & 13:29:53.22  & 47:12:39.5  & $0.09\pm0.02$ & $-0.30\pm0.24$ \\
60*  & 13:29:54.24  & 47:11:23.2  & $0.08\pm0.02$ & $-0.48\pm0.30$ \\
61*  & 13:29:54.32  & 47:11:29.9  & $0.09\pm0.02$ & $-0.26\pm0.24$ \\
62*  & 13:29:54.72  & 47:12:36.6  & $0.10\pm0.02$ & $-0.56\pm0.27$ \\
64*  & 13:29:54.92  & 47:11:33.0  & $0.13\pm0.02$ & $-0.50\pm0.18$ \\
65*  & 13:29:54.95  & 47:09:22.4  & $0.41\pm0.03$ & $-0.30\pm0.09$ \\
67*  & 13:29:55.08  & 47:11:35.0  & $0.13\pm0.02$ & $-0.28\pm0.17$ \\
68*  & 13:29:55.25  & 47:10:46.2  & $0.16\pm0.02$ & $-0.78\pm0.23$ \\
74*  & 13:29:55.52  & 47:12:09.9  & $0.10\pm0.02$ & $-0.49\pm0.23$ \\
75*  & 13:29:55.57  & 47:13:59.8  & $0.18\pm0.02$ & $-0.66\pm0.15$ \\
76*  & 13:29:55.60  & 47:12:03.1  & $0.08\pm0.02$ & $-0.63\pm0.31$ \\
78*  & 13:29:55.69  & 47:11:46.6  & $0.12\pm0.02$ & $<-1.0$ \\
79*  & 13:29:55.86  & 47:11:44.5  & $0.32\pm0.03$ & $-0.27\pm0.10$ \\
84*  & 13:29:57.47  & 47:10:37.1  & $0.13\pm0.02$ & $-0.63\pm0.25$ \\
96*  & 13:30:01.41  & 47:11:57.8  & $0.08\pm0.02$ & $-0.36\pm0.28$ \\
99*  & 13:30:02.03  & 47:09:51.4  & $0.11\pm0.02$ & $-0.33\pm0.22$ \\
104* & 13:30:05.13 &  47:10:35.8  & $9.51\pm0.48$ & $-0.66\pm0.06$ 
\enddata
\end{deluxetable} 

\begin{deluxetable}{lcccc}
\tablecaption{ \label{tab:arp220}
  SNRs in Arp 220 taken from \cite{Parra_etal07}}
\tablehead{ID & R.A. (2000) & Dec. (2000) & S$_{1.4}$ & $\alpha$ \\
 & (hr:min:sec) & ($^{\circ}$:\arcmin:\arcsec) & (mJy) &	}
\startdata
W42 & 15:34:57.2123 & 22:30:11.482 & $0.69\pm0.10$ & $-$0.24 \\
W39* & 15:34:57.2171 & 22:30:11.485 & $3.15\pm0.47$ & $-$1.87 \\
W30* & 15:34:57.2214 & 22:30:11.403 & $1.41\pm0.03$ & $-$1.24 \\
W18* & 15:34:57.2240 & 22:30:11.546 & $1.34\pm0.20$ & $-$0.62 \\
W17 & 15:34:57.2241 & 22:30:11.519 & $0.42\pm0.06$ & $-$0.23 \\
W15* & 15:34:57.2253 & 22:30:11.483 & $2.42\pm0.36$ & $-$0.72 \\
W10 & 15:34:57.2307 & 22:30:11.502 & $0.46\pm0.07$ & $-$0.29 \\
W8 & 15:34:57.2361 & 22:30:11.432 & $0.95\pm0.14$ & $-$1.27 
\enddata
\end{deluxetable} 

\begin{deluxetable}{lcc}
\tablecaption{ \label{tab:lfs}
 SNR LF Parameters}
\tablehead{Galaxy & $\beta$ & $A$\tablenotemark{a} }
\startdata

LMC & $-2.44\pm0.67$ & $11_{-3}^{+5}$ \\
SMC & $-3.44\pm1.47$ & $4.1_{-0.8}^{+1.6}$ \\
IC 10 & $-1.55\pm0.52$ & $2.2_{-1.1}^{+2.2}$ \\
M33 & $-2.25\pm0.17$ & $13_{-3}^{+3}$ \\
NGC 1569 & $-2.47\pm0.46$ & $19_{-4}^{+5}$ \\
NGC 300 & $-3.02\pm0.55$ & $11_{-2}^{+3}$ \\
NGC 4214 & $-2.72\pm1.24$ & $18_{-5}^{+10}$ \\
NGC 2366 & $-2.56\pm1.13$ & $10_{-2}^{+5}$ \\
M82  & $-2.16\pm0.26$ & $670_{-140}^{+180}$ \\
M81 & $-2.63\pm1.18$ & $46_{-15}^{+28}$ \\
NGC 7793 & $-2.17\pm0.70$ & $24_{-13}^{+27}$ \\
NGC 253 & $-2.12\pm0.48$ & $870_{-330}^{+520}$ \\
NGC 4449 & $-1.71\pm0.33$ & $15_{-7}^{+14}$ \\
M83 & $-2.61\pm0.48$ & $98_{-32}^{+47}$ \\
NGC 4736 & $-4.50\pm1.63$ & $100_{-40}^{+70}$ \\
NGC 6946 & $-2.37\pm0.29$ & $140_{-40}^{+50}$ \\
NGC 4258 & $-2.70\pm1.22$ & $270_{-50}^{+130}$ \\
M51 & $-2.36\pm0.25$ & $140_{-50}^{+80}$ \\
Arp 220 & $-3.00\pm1.89$ & $57000_{-19000}^{+39000}$ 
\enddata
\tablenotetext{a}{Calculated assuming $\beta = -2.07$.}
\end{deluxetable}

\end{document}